\documentclass[preprint,12pt,authoryear]{elsarticle}

\usepackage{amssymb}

\usepackage{aas_macros}
\usepackage[utf8]{inputenc}
\usepackage{nicefrac}

\journal{Advances in Space Research}

\begin{document}

\begin{frontmatter}



\title{InfraRed Astronomy Satellite Swarm Interferometry (IRASSI): Overview and Study Results}


\author[l1]{Hendrik Linz}
\author[l2]{Divya Bhatia}
\author[l3]{Luisa Buinhas}
\author[l4]{Matthias Lezius}
\author[l3]{Eloi Ferrer}
\author[l3]{Roger Förstner}
\author[l3]{Kathrin Frankl}
\author[l3]{Mathias Philips-Blum}
\author[l2]{Meiko Steen}
\author[l2]{Ulf Bestmann}
\author[l4]{Wolfgang Hänsel}
\author[l4]{Ronald Holzwarth}
\author[l1]{Oliver Krause}
\author[l3]{Thomas Pany}


\address[l1]{MPIA Heidelberg, Königstuhl 17, 69117 Heidelberg, Germany}
\address[l2]{Institut für Flugführung, TU Braunschweig, Hermann-Blenk-Str. 27, D-38108 Braunschweig, Germany}
\address[l3]{Universität der Bundeswehr München, Werner-Heisenberg-Weg 39, D-85577 Neubiberg, Germany}
\address[l4]{Menlo Systems GmbH, Am Klopferspitz 19a, D-82152 Martinsried, Germany}

\begin{abstract}
The far-infrared (FIR) regime is one of the few wavelength ranges where no astronomical data with sub-arcsecond spatial resolution exist yet. Neither of the medium-term satellite projects like SPICA, Millimetron or OST will resolve this malady. For many research areas, however, information at high spatial and spectral resolution in the FIR, taken from atomic fine-structure lines, from highly excited carbon monoxide (CO) and especially from water lines would open the door for transformative science. These demands call for interferometric concepts. We present here first results of our feasibility study IRASSI (Infrared Astronomy Satellite Swarm Interferometry) for an FIR space interferometer. Extending on the principal concept of the previous study ESPRIT, it features heterodyne interferometry within a swarm of five satellite elements. The satellites can drift in and out within a range of several hundred meters, thereby achieving spatial resolutions of $<$0.1 arcsec over the whole wavelength range of 1--6 THz. Precise knowledge on the baselines will be ensured by metrology methods employing laser-based optical frequency combs, for which preliminary ground-based tests have been designed by members of our study team. We first give a motivation on how the science requirements translate into operational and design parameters for IRASSI. Our consortium has put much emphasis on the navigational aspects of such a free-flying swarm of satellites operating in relatively close vicinity. We hence present work on the formation geometry, the relative dynamics of the swarm, and aspects of our investigation towards attitude estimation. Furthermore, we discuss issues regarding the real-time capability of the autonomous relative positioning system, which is an important aspect for IRASSI where, due to the large raw data rates expected, the interferometric correlation has to be done onboard, in quasi-real-time. We also address questions regarding the spacecraft architecture and how a first thermomechanical model is used to study the effect of thermal perturbations on the spacecraft. This will have implications for the necessary internal calibration of the local tie between the laser metrology and the phase centres of the science signals and will ultimately affect the accuracy of the baseline estimations. 
\end{abstract}


\end{frontmatter}


\section{Introduction}\label{Section:intro}

\begin{figure}[t]
\includegraphics[width=\textwidth]{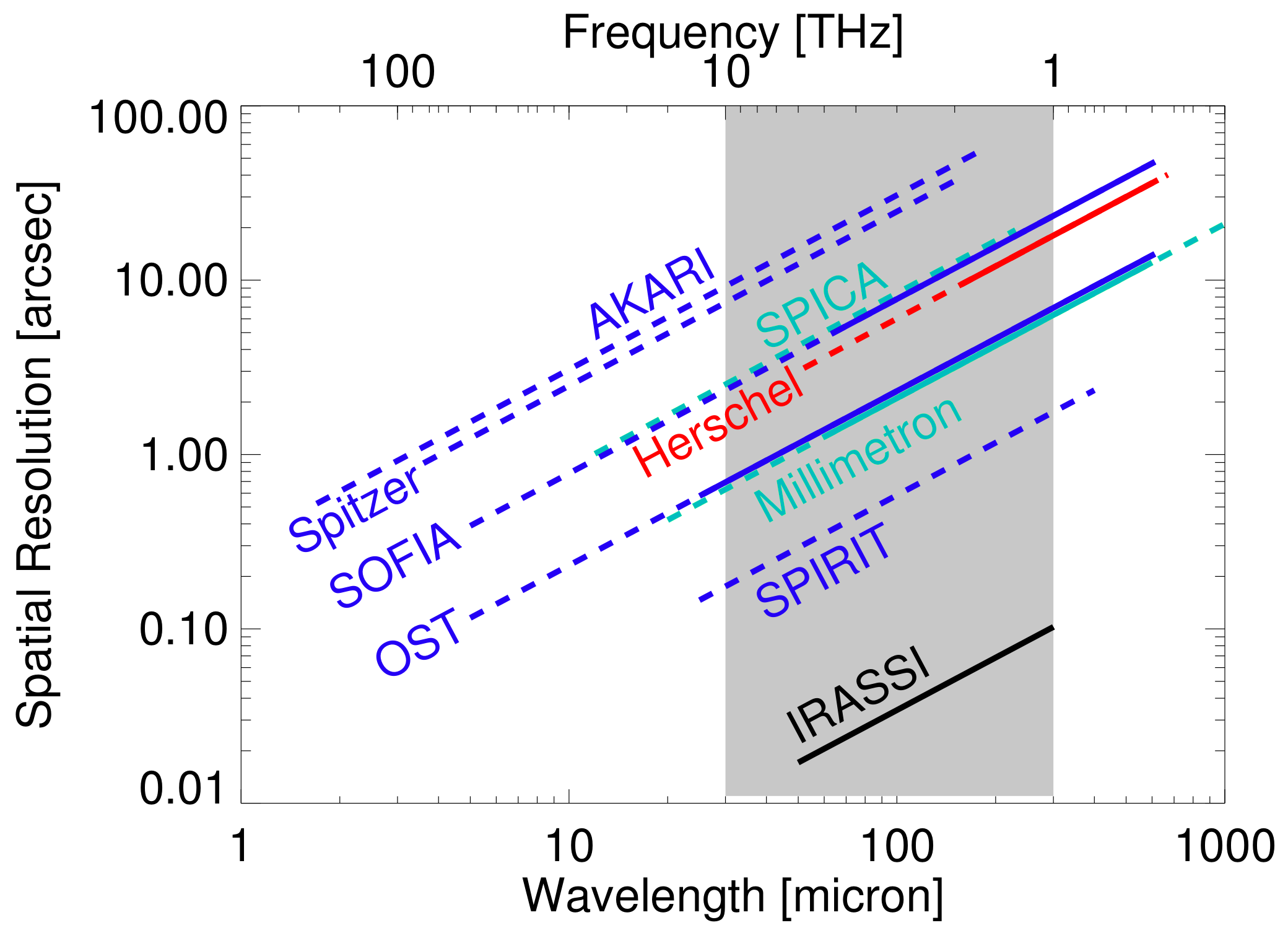}
\caption{Spatial resolution comparison of previous existing FIR missions and for new FIR projects and studies. Solid lines indicate wavelength ranges where high spectral resolution is available. The grey rectangle denotes the FIR range. }\label{Fig:FIR-comparison}
\end{figure}

Many astronomical breakthroughs came with the advance of observing capabilities and the improvement of the achievable spatial resolution. The Hubble satellite has been delivering diffraction-limited data with a spatial resolution of better than 0.1 arcsec in the Optical and UV for 25 years now. In the near-infrared (NIR) one has learned to overcome the disturbing effects of the turbulent atmosphere by means of adaptive optics, enabling observations with spatial resolutions of better than 0.1 arcsec when combined with 8–10 m class telescopes. Even better spatial resolution is achievable in the infrared when combining several telescopes in long-baseline interferometry with facilities like the VLTI or the Keck interferometer. In the radio regime, from early on the adoption of interferometry has been a prime concept.  VLBI techniques achieve sub-milli-arcsec resolution up to frequencies of 86 GHz \citep[and even to 230 GHz; see, e.g.,][]{2008Natur.455...78D}. Common interferometry has spread to ever higher (sub-)millimeter frequencies, with arrays like NOEMA, SMA and especially ALMA as the current state of the art. Hence, we can access the sky at many wavelength regimes with high spatial resolution already. The Far-Infrared (FIR) is a noticable exception. One commonly assigns the wavelength range from 30--300 $\mu$m  (10--1 THz) to the FIR. In this range, the Earth atmosphere is very opaque in general. In particular, in the {\rm interval} from 2--6 THz, the transmission never rises above 1--3 \% even at the best observing sites from the ground, like Dome C in Antarctica \citep{2009P&SS...57.1419S}. Thus, high-flying airborne or space-borne telescopes are imperative to collect astronomical information in the FIR. Currently, the spatial resolution obtainable by the previous and by the currently planned missions is modest. In Fig.~\ref{Fig:FIR-comparison}, we have compiled a plot of the theoretical angular resolution by several past, present and future missions, to depict the situation. A few comments are in order for this plot. The grey area marks the actual FIR range. The lines denote the theoretical spatial resolution (more precisely, the full width at half maximum of the telescope primary beam) for a uniformly illuminated circular reflector with the nominal size\footnote{For OST the optimistic value of 9 m is used \citep[``Concept-1'',][]{2018SPIE10698E..1AS}.} given for these telescopes in the literature. Hence, we neglect that several of these telescopes are not diffraction-limited at all wavelengths, and that the reflectors may be tapered or combined with an undersized secondary mirror, which all diminishes the spatial resolution slightly. We have left out the variety of single-dish balloon missions, which, as far as we know, do not currently feature primary mirrors larger than 3 m. While dashed lines mark the full wavelength range covered by these missions, solid lines show the part of the working range where high spectral resolution (at least approaching a resolution of $10^5$ or better) is available. Currently, the best angular resolution in the FIR has been achieved by Herschel. Also the actively planned single-satellite missions like SPICA 
\citep{2018PASA...35...30R}, the Origins Space Telescope \citep[OST;][]{2018SPIE10698E..1AS}, or Millimetron \citep{2014PhyU...57.1199K} will not push the limit below 1 arcsec (at least not in the 50–300 $\mu$m sub-range of the FIR), although these missions have great merits in being much more sensitive than previous FIR observatories. Only the two-telescope concept previously brought forward for the FIR direct-detection interferometer SPIRIT \citep{2007AdSpR..40..689L} approaches the sub-arcsecond regime. Its pathfinder missions BETTII \citep{2016SPIE.9908E..0SR} -- and hopefully BETTII 2 -- both balloon experiments using two siderostats on a fixed 8-m baseline, also just achieve slightly better than 1 arcsec resolution at 6 THz, but already improve on the Herschel spatial resolution. \\
This status quo suggests to contemplate about future missions that can attain high spatial resolution to fill the gap in the FIR. In the following, we will give a short motivation for selected astrophysical tracers in the FIR. Then we will introduce our concept for a free-flying FIR interferometer, IRASSI, with emphasis on the choice of some of the technical parameters. Furthermore, we mention recent progress in several technological fields that is necessary to make such a mission possible. We then provide an overview of the system architecture and the navigational aspects of our mission study. The paper concludes with an overall summary and outlook on a few key questions that have to be investigated in our upcoming second study phase. 

\section{Scientific Motivation}\label{Section:motivation}

The FIR regime contains some unique tracers of different kinds of astrophysical environments. While other references cover the potential science in the FIR in much more detail \citep[e.g.,][]{2017arXiv170100366R,2018arXiv180400743P,2014PhyU...57.1199K}, we just want to highlight a few aspects regarding the unique tracers we find in the FIR regime. Some of these will also emphasise why high spectral and spatial resolution bring a new dimension to many of the science cases.
\begin{itemize}
    \item The FIR covers many of the most important cooling lines of the interstellar medium. These are predominantly atomic fine-structure lines, like the [CII] transition at 157.7 $\mu$m, the [OI] transitions at 63.2 and 145.5 $\mu$m, the [OIII] lines at 51.8 and 88.4 $\mu$m, and the [NII] lines at 121.9 and 205.2 $\mu$m. Those lines (together with [CI] lines at sub-mm wavelengths) contribute to the energy budget of large parts of a molecular cloud at low and intermediate densities. Hence, in order to understand the complete thermodynamics of the star-formation process in molecular clouds, such tracers are very valuable probes. With velocity-resolved spectral maps in these lines (and complementary ground-based data in lines tracing the dense cold molecular gas), a comparison of the dynamics of the atomic and the molecular gas phase is possible. In particular, [CII]  bears importance also as a star-formation tracer in extragalactic studies, which makes it imperative to understand the [CII] emission in Galactic star-forming regions to be able to extrapolate. We note that [$^{12}$CII] can attain high optical depths \citep{2013A&A...550A..57O}. A triple of [$^{13}$CII] hyperfine-structure lines exists as well around 1.9 THz, with the closest component separated by just $\sim 11$ km/s from the [$^{12}$CII] line. These isotope lines can be used to assess the optical depth of the main line; obviously, high spectral resolution is mandatory here to disentangle the lines.  
    \item A strong motivation for observations from space is to observe many transitions of the water molecule 
H$_2$O, a task that is notoriously difficult from the ground.  Water is relatively easily removed from grain surfaces and hence quite abundant in warmed-up regions, and also in the presence of shocks. Furthermore, in circumstellar disks, the direct detection of water gas in a spatially resolved fashion gives precious information about the water budget in such disks, and would more directly indicate where the (water) snow line is located. This would give very valuable input to gauge planet-formation scenarios. Previous studies showed a low Herschel/HIFI detection rate of H$_2$O ground-state transitions at longer wavelengths for a small sample of TTauri disks \citep{2017ApJ...842...98D}. However, two studies for T Tauri disks and for Herbig Ae disks by \citet{2016ApJ...827..113N,2017ApJ...836..118N} present modelling of higher-excited water lines that better trace the central disk part and are not easily excited in the thinner disk atmosphere. Notsu et al.~did not consider the possibility that one could spatially resolve H$_2$O emission from inside the water-snow line in the FIR. But given the maximum baseline lengths for IRASSI, an angular resolution of around 20 milli-arcsec can be achieved for some of such lines, which corresponds to roughly 3 au at the distance of several nearby low-mass star-forming regions (140 pc). Hence, for protoplanetary disks moderately close to Earth, IRASSI could separate the water-gas emission from inside the water-snow line from emission arising from the disk upper layers further out.
    \item Molecular gas is predominantly consisting of H$_2$ molecules which, however, do not possess a permanent dipole moment. Only quite hot H$_2$ gas can be seen in quadrupole rotational and roto-vibrational transitions in the near- and mid-infrared. The singly deuterated isotopologue hydrogen deuteride (HD) has a rotational spectrum, and may be used as a unique proxy to the bulk reservoir of molecular gas in a variety of contexts. Its usefulness as a disk mass tracer, cum grano salis, has recently been shown in works based on Herschel/PACS spectra \citep{2013Natur.493..644B,2016ApJ...831..167M}. The first two rotational transitions lie at 112.1 $\mu$m and 56.2 $\mu$m. Employing HD as a tool is hence a unique capability of the FIR range.
    \item The FIR contains a ladder of high-lying rotational transitions for the abundant molecule CO, with upper level energies E$_{\rm up}$/k$_{\rm B}$ ranging from 249 K up to several thousands of Kelvin. Such tracers probe the warm and hot molecular gas, for instance close to low- and high-mass protostars. Having velocity-resolved data at hand is pivotal for disentangling gas directly heated by the central protostar and gas heated and compressed due to the presence of shocks.
    \item Regarding the continuum sensitivity at the high frequencies of the FIR, a heterodyne interferometer might be at a disadvantage compared to direct-detection interferometers and bolometers with their wide bandwidths. Fortunately, in the FIR we catch the dust continuum emission from cold and embedded objects at the peak of their spectral energy distributions. Combining this with the fact that the development of sensitive and wide bandwidth heterodyne receivers and backends steadily progresses, it will be possible to deliver high-resolution maps of the FIR continuum for many dust-dominated astrophysical object classes. One very promising aspect in connection with the strong FIR continuum is the detection of gas lines in absorption against the continuum. High spectral resolution can reveal line asymmetries which hint at infall motions. With the instrument GREAT \citep{2012A&A...542L...1H} aboard the SOFIA facility it has been possible to exploit a NH$_3$ line at around 1.81 THz for that purpose \citep{2012A&A...542L..15W,2016A&A...585A.149W}. With a beam size of 16$''$, it was possible to trace inward motions with SOFIA on larger scales, averaged over the extended envelope of a few high-mass protostars. With a facility like IRASSI, we can pinpoint on a hundred times finer scale how the true infall rate is towards the actual protostar.
\end{itemize}
In conclusion, a FIR interferometer like IRASSI can exploit the unique line tracers that this wavelength range provides. It will be particularly fruitful for the research of structure of the planet-forming regions in circumstellar disks, as well as for gas dynamics in the close vicinity around low- and high-mass protostars. But also studies of giant molecular clouds in external galaxies, and towards AGNe, will greatly profit from the capabilities of such a facility.

\section{Principle Concept for IRASSI}\label{Section:concept}

We present here the concept of a space-borne FIR interferometer. We gave it the name IRASSI (Infrared Astronomy Satellite Swarm Interferometry). Currently, this is an extended study, funded by the German Space Agency DLR, for which three German research institutes and an industry partner have joined forces. The initial ideas for IRASSI are strongly rooted in the previous study ESPRIT \citep{2008SPIE.7013E..2RW} that had been published 11 years ago, and which was one contribution to the FIRI initiative for the ESA Cosmic Visions 2015--2025 \citep{2009ExA....23..245H}. IRASSI presents also an extension to previous studies since it actively addresses questions of navigation, swarm dynamics and internal calibration. The central ideas of such a facility are: \\[-8mm]
\begin{itemize}
    \item swarm operation in a Halo orbit around Lagrange point L2 \\[-8mm]
    \item freely drifting satellite elements to sample the (u,v) plane \\[-8mm]
    \item satellite distances do not have to be controlled but just accurately measured and communicated to the correlator \\[-8mm]
    \item inter-satellite metrology via laser-based optical frequency combs \\[-8mm]
    \item employing heterodyne interferometry, detecting electromagnetic waves with amplitude and phase information
\end{itemize}
Table \ref{Table:requirements} summarises the main requirements for several of the key parameters of IRASSI. In the following, we will elaborate on the motivation and reasoning for some of the choices we have made. 
\begin{table}[t]
\begin{tabular}{l|l}
Parameter                                           &  Required Value \\
\hline
\hline
Number of telescopes                            & 5 \\
\hline
Number of immediate baselines               &  10 \\
\hline
Size of telescope mirrors                        &  3.5 m (primary mirror) \\
\hline
Spacecraft configuration                         &  Free-flying in 3D,  \\
                                                         &  changed by initiated passive drifts  \\
\hline
Length range of projected baselines         &  7-10 m (min) to 850 m (max) \\
\hline
Essential wavelength range                     &  50 to 300 $\mu$m \\
\hline
Essential frequency range                       &  1 to 6 THz \\
\hline
Frequency bandwidth                             & 2 GHz (min. requirement); \\
                                                         & 16 GHz dual polarisation (goal) \\
\hline                                                         
Velocity resolution                                 &  $<$1 km/s \\
\hline
Useful Field of view                               & 3$''$ to 18$''$ (frequency-dependent) \\
\hline
Angular resolution                                 & 0.1$''$ (at 300 $\mu$m) \\
\hline
Telescope pointing accuracy                    & 0.4$''$ (requirement), 0.2$''$ (goal) \\
\hline
Accuracy of baseline metrology                & $< 5 \, \mu$m (3D) \\
\hline
Temperatures                                       & $\approx 80$ K (main dish, passive cooling); \\
                                                         & 4 K (mixer, active cooling) \\
\hline
\end{tabular}
\caption{Basic parameters for the IRASSI concept of a FIR space interferometer.}\label{Table:requirements}
\end{table}
  
\subsection{Number of telescopes}\label{Subsection:tel-num}
\begin{figure}[t]
\includegraphics[width=\textwidth]{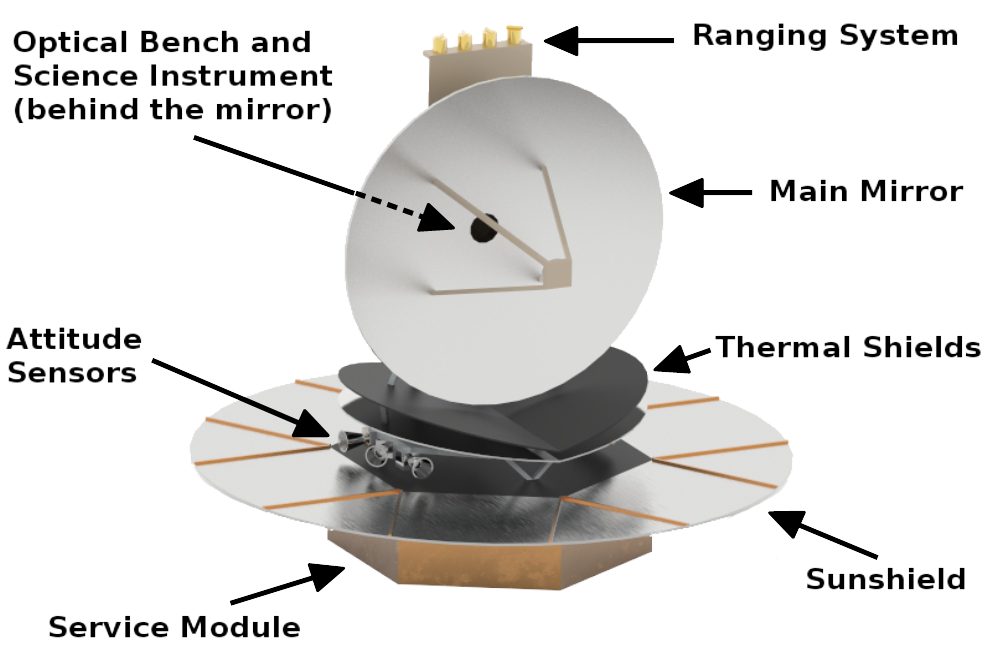}
\caption{A sketch of the principle design for an IRASSI satellite element. Taken from \citet{Ferrer-Gil_2016}. The drawing is roughly to scale. The ranging system consists of four laser portals for inter-satellite metrology and data transfer and is on the cold side of the sun and heat shields. Among all the attitude sensors, just the star tracker cameras are explicitly visible here, while gyroscopes will be mounted in the interior of the satellites.}\label{Fig:IRASSI-sketch}
\end{figure}
Several considerations of different nature influence this decision. On the one hand, more telescopes offer more immediate coverage of the (u,v) plane and of the related spatial frequencies. For a number of N interferometer elements, there are N (N-1) / 2 immediate baselines, hence, with increasing number of elements, the number of baselines increases faster than linearly. However, with more and more interferometer elements, the complexity of the mission rises drastically. The choice of the telescope number of course affects most sub-systems, from the demands on the total data transfer between all the elements up to the complexity of the laser metrology system. As another boundary condition, one also has to keep in mind, that only a few interferometer elements might be put into space simultaneously with one launch. We had done the simulations for launch and transfer to L2 assuming one Ariane 5ECA\footnote{One has to mention that the Ariane 5 model series is due to be discontinued in the future, after the JWST launch.} launch for 5 satellite elements which seems feasible \citep{Buinhas_2016}. Certainly, a launcher like the ``Falcon Heavy'' would be an even more powerful alternative. After the update on the parameters for the Ariane 6-64 in March 2018\footnote{Lagier, R., et al., "ARIANE6", User’s Manual, Issue 1, Rev. 0 (March 2018), http://www.arianespace.com/wpcontent/uploads/2018/04/Mua-6\_Issue-1\_Revision-0\_March-2018.pdf .}, one has to check if also this launcher would be capable to master this task. 
One can start from a different point of view and ask which number of telescopes is essential to achieve certain goals and to recover important quantities. In principle, even with two telescopes (and a meticulously planned observing strategy) it might be possible to cover considerable parts of the (u,v) plane. However, with two telescopes it is not possible to recover the so-called closure quantities \citep{2001isra.book.....T} which provide interferometric information that is not affected by certain measurement errors.\\
\underline{Closure Phases:} In order to assess this quantity, at least three immediate baselines (and thus three telescopes) are necessary. Though the information is now reduced from three Fourier phases to one closure phase quantity, it is not affected by additive phase errors anymore. In general, for N interferometer elements, (N-1) (N-2) / 2 independent closure phases can be constructed \citep[][p. 400]{2001isra.book.....T}. Closure phases give important constraints on the geometry of the intensity distribution to be reconstructed. While unresolved point sources result in closure phases of $0^\circ$ or $180^\circ$, any deviation from these values indicates a deviation from point-symmetry for the intensity distribution \citep{2007NewAR..51..604M}.\\
\underline{Closure Amplitudes:} The closure amplitude is an independent quantity calculated for a combination of four stations.  For this quantity, amplitude errors cancel out similar to the individual phase errors in the closure phase. The closure amplitude is thus independent of all station-dependent amplitude errors, such as pointing errors or dish deformation. It can be used to compensate for detector gain fluctuations and changing antenna efficiencies. It does not eliminate baseline-dependent errors, though. For a set of N interferometer elements, there are [N (N-3)] / 2 independent closure amplitudes \citep[][p. 401]{2001isra.book.....T}. For an unresolved point source, the closure amplitudes should be unity.
While closure phases have been used already for a long while in both radio and infrared interferometry, closure amplitudes are not so well known yet, although they are used implicitly in radio interferometry data models. One example where closure amplitudes have been used explicitly is the work of \citet{2001AJ....121.2610D} on VLBI observations of the Galactic center at 86 GHz.
Taking all this into account we opt for a five-element design as a baseline requirement for IRASSI. It allows to retrieve precious closure phase and closure amplitude information. Even if one of the five elements is compromised, still a reduced number of closure quantities can be recovered. With five interferometric elements there are: 10 immediate baselines = 10 immediate Fourier phases and amplitudes, 6 independent closure phases, and 5 independent closure amplitudes.

\subsection{Baseline lengths – Baseline accuracy}\label{Subsection:baselines}
One of the basic goals of an interferometer like IRASSI would be to deliver sub-arcsecond spatial resolution over the whole FIR frequency range, from 1 to 6 THz. We set as a requirement for IRASSI to achieve 0.1 arcsec angular resolution even at the lowest frequency of 1 THz $\widehat{=} \, 300 \mu$m. In order to estimate the maximum baseline lengths necessary in order to deliver such an angular resolution value, we start with a simple estimate. Assuming that the (u,v)-plane is sufficiently filled, the uniformly weighted sampling function out to a maximum circle (u,v)-radius of $r_{\rm max}/\lambda = u_{\rm max}$ (with $r_{\rm max}$ as the maximum baseline and $\lambda$ as the wavelength) will give rise to a synthesised beam in form of a Bsinc function whose FWHM = $0.705 \, u^{-1}_{\rm max}$ \citep[][p.389]{2001isra.book.....T}. To achieve FWHM=0.1 arcsec at 300 $\mu$m calls for baseline lengths up to $\sim 425$ m according to this relation. Uniform weighting delivers the best spatial resolution, but comes with a relatively high sidelobe level. There will be many situations where a trade-off has to be made between spatial resolution and noise performance and, consequently, when natural weighting or certain kinds of (u,v)-tapering will be necessary in post-processing. In order to ensure also in such situations the 0.1-arcsec resolution, we introduce a factor of 2 for the maximum baselines. Hence, IRASSI shall be operational out to maximum baseline lengths of around 850 m as a firm requirement. 
Regarding the necessary baseline length accuracy, we mention again that we do not need to control these distances. We just need to accurately measure them in order to inform the correlator about the geometric delays to be taken into account. Observations at the shortest wavelengths naturally have the highest demands. A minimum requirement is that the knowledge about the baseline lengths should be better than $\lambda_{\rm min}/10$ in order not to accumulate too large phase uncertainties. Hence, the baseline length measurement accuracy (in 3D) should be better than 5 $\mu$m for IRASSI where $\lambda_{\rm min} = 50 \mu$m. In such a setting, we then have to deal with phase uncertainties of several tens of degrees (for the shortest operational wavelengths of IRASSI), as it commonly occurs in real ground-based interferometry at elevated frequencies due to atmospheric decorrelation effects. We enter a different level of requirements, though, if we demand that baselines are to be calibrated (instantaneously in our case) in the same sense as interferometer element distances are being calibrated on ground. There, by applying calibrator measurement series on unresolved reference sources, one can attempt to push the uncertainties of the baseline lengths down to $\lambda$/360, hence approaching phase uncertainties as small as $1^\circ$ for these reference phases \citep[][p. 93]{2001isra.book.....T}. This would put much more stringent requirements for the metrology systems. We will briefly come back to this issue in Section \ref{Subsection:frequency-combs}.

\subsection{Telescope diameter – Pointing accuracy}\label{Subsection:pointing-accuracy}
In subsection \ref{Subsection:sensitivity} we recapitulate how the main reflector size of the individual satellite elements is related to the final sensitivity of the interferometer; in particular, that a scale-down of the diameter would have to be compensated by a disproportionate increase of the number of antennas. Therefore, for the IRASSI study, we consider a small swarm with large but still practical telescope diameters.
We assume that the main reflector dishes of the interferometer elements have a diameter of 3.5~m. This choice is of course closely based on the FIR/sub-mm satellite Herschel \citep{2010A&A...518L...1P} which featured the same telescope diameter. Such mirrors still fit into the fairings of currently available large rocket launchers. The relatively large size results in ample collecting area per IRASSI element. This is a crucial prerequisite to conduct sensitive spectral line studies (cf. subsection \ref{Subsection:sensitivity}), which is of course the main working area for an interferometer like IRASSI. \\
The choice of the mirror size has implications for instance for the necessary pointing accuracy of the individual antenna elements. Keep in mind that for a uniformly illuminated antenna with circular aperture of radius r the resulting power pattern (the “primary beam”) has a half-power beam width HPBW = $0.514 \lambda$/r and a first null at $0.610 \lambda$/r \citep[][p.389]{2001isra.book.....T}. (A realistic choice for such antennas will deviate from uniform illumination in order to lower the sidelobe levels and to avoid standing-wave issues, e.g., by introducing offset or off-axis designs.) Nevertheless, the flanks of the power pattern for most designs are usually quite steep. It is obvious that antenna pointing errors (especially the relative pointing error RPE, see below) will (temporarily) displace the maximum of the reception pattern from the observing direction (i.e., the phase centre). Our demands on the pointing accuracy will depend on the short end of the wavelength operational range of IRASSI. With $\lambda_{\rm min} = 50 \,\mu$m, the HPBW is 3 arcsec for a uniformly illuminated circular antenna. Modified designs for the main and secondary mirrors will slightly widen the main lobe by approximately a factor of 1.1--1.4. Assuming a factor of 1.33 compared to the uniform illumination HPBW, we set the minimum requirement for the pointing accuracy to HPBW$_{\rm eff}$(50 $\mu$m)/10 = 0.4 arcsec. However, in order to ensure high-quality interferometric data, it is suggested to pursue even higher accuracies. \citet{1999ASPC..180...37N}, for instance, advocates a pointing accuracy of HPBW/20. With this performance, a source located in the centre of the antenna beam will suffer only negligible intensity variations since the reception pattern at HPBW/20 is still around 0.995. The fractional intensity variation for a source located at the half power point of the pattern, however, will already be $\approx 0.87$, even if the strict limit of HPBW/20 is enforced for the pointing accuracy. One can see that the accuracy of the outer part of the image will be noticeably reduced. Therefore, HPBW$_{\rm eff}$(50 $\mu$m)/20 = 0.2 arcsec is a very desirable goal for the IRASSI pointing accuracy. This goal is desirable for both, absolute and relative pointing errors. Compared to the values achieved with the HERSCHEL satellite of 0.9 arcsec (see Sect.~\ref{Subsection:ADCS}), this calls for a clear improvement of a factor 3--4, but is not unrealistic (cf.~Sect.~\ref{Subsection:ADCS}). 

\section{Technical progress on several fronts}\label{Section:technical-progress}

\subsection{Data rates and necessity for on-board correlation}\label{Subsection:data-rates}
A principle question is how much raw data are being produced, and whether this amount can be downlinked to Earth for a posteriori correlation. The raw data rates can be approximated with this formula \citep[cf.][]{2016ExA....41..271R}:
\begin{equation}\label{Equation:raw-dr}
D_{\rm obs}  =  2  \Delta \nu  N_{\rm pol}  N_{\rm bits}  N_{\rm sat}  \,\,\,  [{\rm bit/s}]					
\end{equation}
For IRASSI, $N_{\rm sat} = 5$. Let us assume that just two distinct polarisation products will be evaluated ($N_{\rm pol} = 2$) and not the full four Stokes parameters. Let us further assume a bit-rounding level for the sampler/digitiser of $N_{\rm bits} = 2$. Regarding the spectral bandwidth $\Delta \nu$ [in Hz] to be covered, we have to keep in mind the boundary condition that IRASSI should operate at high frequencies up to 6 THz. To cover a moderate velocity range of $\pm50$ km/s around a spectral line, already 2 GHz of bandwidth are necessary at this highest frequency. If IRASSI will operate in a communication mode similar to the Herschel satellite, we could assume 20 hours of autonomous operation (see also Sect. \ref{Sect:Formation-flying}), and a maximum window of 4 hours of data transfer per operational day. With the chosen parameters, IRASSI will produce an amount of raw data of 5.76e+15 bits every 20 hours. To put this into perspective, we have to compare it to the data that could realistically be downlinked within the communication window. We give two examples.  \\
(1) The EUCLID satellite (to be operated in L2, launch $\sim$2021) will employ a new K-band (26 GHz) radio telemetry system for the main data transfer. This will enable to downlink the total amount of 850 Gbit of (compressed) data per operational day within a 4-hour communication window, whereby the maximum downlink rate is 74 Mbit/s \citep{2014SPIE.9143E..0HL}. The severe mismatch between the 8.5e+11 bit/day for EUCLID and the 5.76e+15 bit/day for IRASSI is obvious. \\
(2) Looking more into the future, studies have been undertaken to estimate the data transfer rate using laser links directly to Earth, employing the large telescopes of modern Cherenkov telescope arrays as collecting bowls \citep{2015IPhoJ...700203C}. This work estimates effective data rates of around 700 Mbit/s from an L2-orbit to Earth using this method. The possible amount of data to be transferred within a 4-hours communication window is then 1.01e+13 bit/day.
One can conclude that such techniques are not sufficient to enable a downlink of the IRASSI raw data to Earth. Even if continuous data downlink capabilities, operating simultaneously to the actual interferometric measurements, could be implemented, and raw data compression factors of 10 could be achieved, it would still take almost 9.5 days to downlink the raw data of one day. (This is just the exercise for the minimum bandwidth of 2 GHz. Remember that there is the ambition to correlate larger bandwidths of 8 or even 16 GHz.)
As a consistency check, we also evaluate the data rates after correlation \citep[cf.][]{2016ExA....41..271R}:
\begin{equation}\label{Equation:correlated-dr}
 D_{\rm cor}  =  2  (N_{\rm pol}   N_{\rm sat} )^2  N_{\rm bits}  N_{\rm bins}  /  T  \,\,\,  [{\rm bit/s}]
 \end{equation}
For producing these data, one should employ a higher-quality digitisation, e.g., $N_{\rm bits} = 16$. With $N_{\rm bins} = 1024$ spectral bins, and with a typical correlator averaging time of T=1 s, we arrive at a data rate of 3.28 Mbit/s. Integrated over 20 hours of operation, this amounts to 2.36e+11 bits, which can easily be downlinked within 3200 s when the 74 Mbit/s download rate of a EUCLID-like system is available. A more detailed calculation for the download budget for IRASSI, including realistic parameters for the transmission system (transmitter antenna efficiency, transmission path, atmospheric attenuation, etc.) can be found in our related publication \citep{Ferrer-Gil_2016}.  

\subsection{Distributed correlator – power consumption}\label{Subsection:distributed-correlator} 
Similar to the ESPRIT study, we therefore also adopt the principle idea of on-board correlation. Such concepts have been discussed for a while now for potential very-low-frequency radio interferometers in space, both with a central correlator unit \citep[e.g.,][]{2005RaSc...40.4004O} or with a distributed correlator \citep[e.g.,][]{2016ExA....41..271R}. For IRASSI, we prefer a setup with a frequency-distributed correlator where all five satellite elements act in a symmetric fashion, and each instance of the distributed correlator correlates 1/5th of the entire spectral bandwidth. Such a symmetric configuration avoids the single-point failure of one central correlator. Even in case one of the five correlator elements is compromised, still 80\% of the full bandwidth could thus be evaluated.
The numbers for the raw data rates elaborated on in the previous subsection have also implications for the inter-satellite data communication for such a distributed correlator concept. In the 2-GHz-minimum-bandwidth example, every satellite produces 1.15e+15 bits of raw data per 20 hours. 4/5 of that have to be transmitted to the other 4 satellite elements. Hence, every satellite also receives 4/5th of that one-satellite budget = 0.92e+15 bits within 20 hours  (i.e., data volume for transmission = data volume for reception). This in turn means that a data exchange rate of  12.8 Gbit/s has to be achieved per satellite. Wider spectral bandwidths of 8 or 16 GHz will push these demands even higher. This calls for a laser data communication system. We currently discuss whether the metrology laser system can be co-used for this purpose. We conceive that the correlation is to be done in quasi-real time. Storage space for the huge amount of raw data would fill up quickly. Hence, the capacity for buffering raw data would be perhaps just in the order of minutes. (For the above example of 2 GHz bandwidth, every satellite will produce 120 Giga-Byte per minute.) We have not estimated yet the speed of correlation, but expect that also to be following real-time. Note that the correlation period can be extended into the communication time window which is several hours long.
The necessity to correlate such large amounts of raw data raises the question about the power consumption of the correlator modules. Fortunately, technological developments in recent years have been quite promising. The previous generation of GPU- or FPGA-based correlator modules still showed relatively high figures-of-merit for the energy per complex multiply-accumulate (CMAC) operation (typically around 1.0e-9 Joule). A new generation of application-specific integrated circuits (ASICs) based on CMOS technology promises much more energy-efficient computing. In the recent study by \citet{2016JAI.....550002D} such modules are introduced which (for an optimal combination of telescope numbers and bandwidth) can reach 1.8e-12 Joule per CMAC operation for the actual cross-correlation. Even when taking into account further energy demands of such components on a system level, the authors estimate that the total figure-of-merit is still below 1.0e-11 Joule, hence showing a factor of 100 improvement compared to the previous generation of correlator modules. One such ASIC device dissipates 1.824 W (for the actual CMAC operations), and shows a power consumption on the system level (including board-level I/O, power supplies, and controls) on the order of 5 W. One has to admit that the design presented there is optimised for large telescope numbers ($\ge 128$) and small bandwidth (a few MHz) per module. But these two parameters work in a kind of trade-off. Given the small number of telescopes for IRASSI, such a module could correlate already a bandwidth of several GHz, sufficient for the correlator setup per satellite in our frequency-distributed correlator frame work.

\subsection{Mixer performance at THz frequencies}\label{Subsection:mixers}
Much progress has been achieved regarding the technology that enables the first steps in the detection chain for Tera-Hertz heterodyne observations., i.e., receivers, local oscillators and mixers. Since the ESPRIT study \citep{2008SPIE.7013E..2RW} we have seen the successful operation of the HIFI instrument (up to 1.9 THz) onboard of the Herschel Space Observatory between 2009-2013, and the ongoing operation of the GREAT/upGREAT instrument (up to 4.7 THz) associated with the SOFIA airborne observatory. A common theme for development in recent years is the combination of Hot Electron Bolometer (HEB) mixers \citep[e.g.,][]{2016SuScT..29b3001S} and Quantum Cascade Lasers (QCL) as local oscillators for the heterodyning \citep[e.g.,][]{2013JIMTW..34..325H,2015OExpr..23.5167V}.
One important quantity to compare the performance of heterodyne systems to is the quantum noise limit (TQL = h $\nu$ / k$_{\rm B} \sim 48$ K/THz for single-sideband operation). A figure-of-merit often invoked is 10 $\times$ TQL, an important border that has to be vanquished in order to stay competitive with incoherent detector systems. More and more experimental THz receiver setups including HEB mixers reach or even transcend this border \citep[e.g.,][]{2013ApPhL.102a1123K}. Currently, values down to 5 $\times$ TQL have been achieved \citep{2018ITTST...8..365K}. A recommendation in the FIR Roadmap prepared for ESA \citep{2017arXiv170100366R} is to aim for a performance better than 3 $\times$ TQL, in order to achieve many of the ambitious science goals collected in that document.  Given the current rate of improvements, such a performance will be in reach in the next decade. We note again, that the part of the signal chain will need active cooling to suppress thermal noise in these sensitive devices. New superconducting materials are being tested for inclusion in the HEB mixers. For instance, MgB2 promises a good combination of low noise, wide noise bandwidth, and the possibility to operate such mixers at slightly elevated temperatures beyond 5-20 K \citep{2017ApPhL.110c2601N}. On the other hand, new setups promise to place the mixer and heterodyne in the same cryostat \citep{2017R&QE...60..518S}, which will be beneficial for further miniaturisation of such structural components. 

\section{Focus points of our study: Navigation, system architecture, baseline metrology, attitude estimation, sensitivity}

IRASSI is undoubtedly a complex and ambitious mission concept. The focus of our own work in the IRASSI team, besides the advancement of accurate metrology via laser frequency combs, is on navigational aspects of such a mission, and to devise a realistic view on the system architecture of such satellites. We give here just short summaries of our work and refer to our original publications for detailed descriptions.

\subsection{Formation flying, relative dynamics, and modes of operation}\label{Sect:Formation-flying}
\begin{figure}[t]
\begin{center}
\includegraphics[width=6cm]{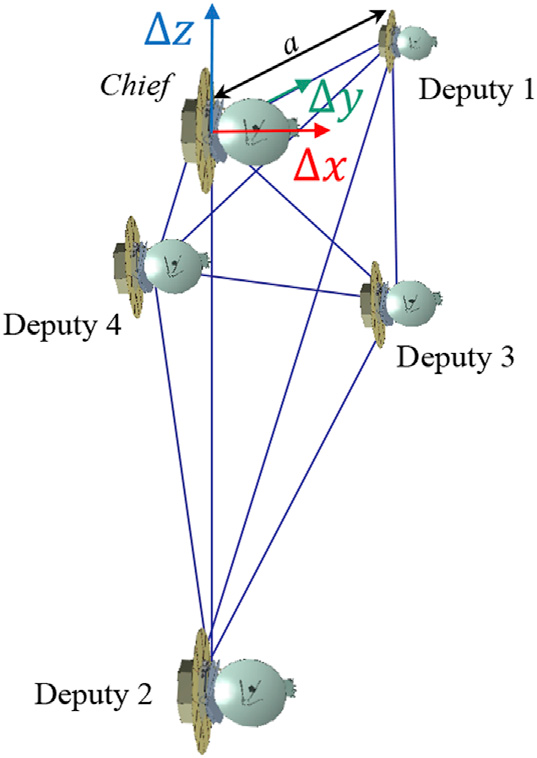}
\end{center}
\caption{Sketch of the nominal formation geometry \citep[adapted from][]{Buinhas_2018}.  In this general case, the formation will not be co-planar, but in a true 3D configuration (see text).}\label{Fig:formation}
\end{figure}

\begin{figure}[t]
\begin{center}
\includegraphics[width=\textwidth]{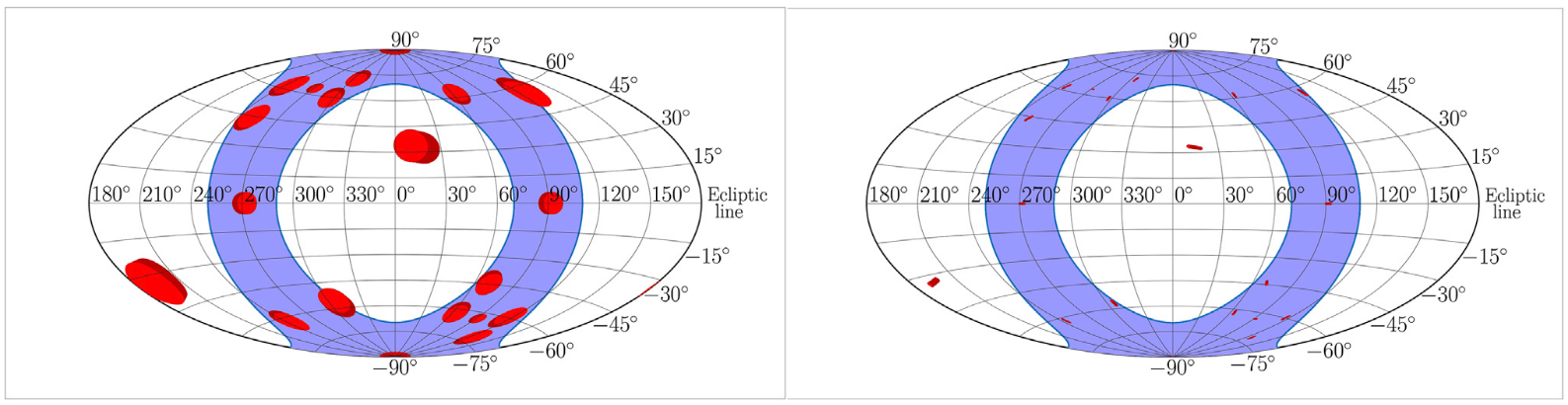}
\end{center}
\caption{Instantaneous sky coverage (blue) and exclusion regions (red) for two sizes of the initial satellite configuration (see text above, and Fig.~\ref{Fig:formation}). Left: The small configuration scenario, where the length parameter a = 40 m. Right: The large configuration scenario, where the length parameter a = 400 m.}\label{Fig:exclusion}
\end{figure}

A full simulation has been performed in \citet{Buinhas_2016} for the launch and transfer of the 5-satellite swarm to the Lagrange point L2, where the satellites are to be injected into a Halo orbit, using the General Mission Analysis Tool (GMAT). Furthermore, a Delta-v budget for necessary accelerations over the mission timeline has been estimated, which includes contributions for the initial transfer to L2, station keeping (to remain in a quasi-Halo orbit around L2), and for routine operations (formation reconfiguration, baseline change initiation, fine target acquisition).  The total number is estimated to be around 300 m/s per spacecraft \citep{Buinhas_2017}. This has implications of course for the choice of the propulsion systems, whereby different requirements are valid for different tasks. Especially the more frequent tasks requiring high control precision take more than 80\% of this budget. Different technologies have been identified which can provide the thrust level required by IRASSI for such tasks. These include pulsed plasma thrusters (PPT), field emission electrostatic propulsion (FEEP), colloidal thrusters, and cold gas thrusters. PPTs might have issues since pulsed operations may introduce disturbances; this needs further investigation. Cold gas thrusters are probably discarded since they require a large amount of fuel. Due to the experimental nature of FEEPs, currently colloidal thrusters seem to be the prime candidates for a secondary propulsion system. The (fine) target acquisition at the beginning of a measurement can call for a quite extended time span of around one hour or more, especially if one decides to do this via micro thrusters and not via reaction wheels. This includes time for internal pre-calibration, laser re-acquisition, and time to allow flexible motions to subside after reconfiguration and re-pointing.\\
The distance to the Sun-Earth L2 location is such that realtime communications and availability are limited and that the spacecraft ought to be embedded with a high degree of autonomy to maneuver during the operational phase throughout the mission, as well as for carrying out vital housekeeping tasks, such as avoiding collisions. We encounter several challenges: a stringent position accuracy requirement, the real-time capability of the positioning algorithm and the deviation of the measured satellite distances between some ranging system and the desired distance between the telescope reference points. In \citet{Frankl_2017}, we approach these challenges by developing and testing two autonomous relative positioning algorithms, based on a geometric snapshot approach. The goal of these algorithms is to determine the coordinates of the telescope reference positions, treating the full 3D uncertainties of the reference points. One promising algorithm transforms the distance measurements between the ranging systems to the telescope reference positions using attitude and angular measurements, while subsequently an optimisation problem is formulated taking into account only the transferred distance measurements. 
Follow-up work in this regard has been presented by us in \citet{Philips-Blum_2018}. The results of this paper show that the required baseline accuracy between the satellites of 5 $\mu$m \emph{cannot} easily be achieved by the basic snapshot method with the currently assumed measurement accuracies of around 1 $\mu$m. The geometric snapshot method is time-consuming in that it requires several iterations to achieve convergence. Thus, a filter-based approach will be considered in the future. Since filter-based approaches incorporate a process model, the influence of acceleration forces on each satellite was investigated. It could be shown that the influence is major such that different accelerations need to be taken into account for any satellite in general. In future work, a filter-based approach considering such accelerations will be investigated.\\
We also simulated realistic scenarios for the swarm configurations \citep{Buinhas_2018}. We found that if arranged in a strictly planar configuration, the relative positioning determination procedure is hampered. We therefore treat the general case of non-coplanar geometries, where the satellites are staggered in 3D in a kind of pyramid configuration (see Fig.~\ref{Fig:formation}). Here, the (x,y,z) coordinates are assumed to be at the initial formation locations (0,0,0), (0,a,0), (0,0,-2a),(a/4,a/2,-a), and (-a/4,a/3,-2/3 a), and subsequently drift motions are initiated and followed by the dynamics computation.   
For these simulations, we take the concrete geometry of the satellite, including the size of the sun-shield, and the location of the laser metrology terminals into account for predicting the sky coverage, and the location of exclusion regions. The latter occur due to at least two instances: (1) the fact that the sun-shield creates an exclusion cone for the laser beams when looking “down” along its own host satellite. When no laser metrology is possible to one of the other satellites such a setup has to be avoided, since in such a case the correlation of the science data would be compromised. (2) Furthermore, when satellites come close, shadowing of one satellite by another has to be avoided, obviously, to not block the FIR science signal from reaching the receiver. The resulting effects are depicted in Fig.~\ref{Fig:exclusion}. The blue regions show the immediate sky coverage for a specific epoch. As the IRASSI swarm revolves around the Sun within one year, basically the whole sky can be covered eventually. But for a concrete epoch, targets located in the exclusion regions have to be avoided. A suitable target allocation strategy must therefore ensure that the physical disposition of the formation does not inhibit the observations throughout the entire observations period. \\
An important capability of IRASSI will be to probe a variety of spatial frequencies, and to eventually probe compact emission with resolution elements of $<$ 0.1 arcsec in the far-infrared. IRASSI shall be able to integrate and take interferometric data while gradually changing the baselines. The above mentioned case of 20-hour drifts are for such applications for which the combination of good (u,v) coverage and high sensitivity is mandatory, for instance when a high-fidelity image reconstruction is being attempted for weak line sources. In principle, other modes of operation are possible. Larger surveys of moderately bright sources could utilise a strategy where the swarm is kept in one configuration without relative drifts, observing several possible targets in a row, and changing to different distinct configurations and repeating the observational series consecutively. The (u,v) coverage will be more sparse, but may still allow an image reconstruction. Then, of course, the data acquisition is spread over many operational days. This implies that the final data post processing on ground can just start when all data have been gathered, and the scheduling for such series should be done in a way that all the necessary configurations can be met before some targets leave their visibility patch for several weeks or even months. Finally, for strong sources, also a snapshot mode can be envisioned where important information can already be obtained from relatively short observations on the order of one or a few hours in one configuration. A full image reconstruction may not be possible. But crucial interferometric parameters like closure phases (see Sect.~\ref{Subsection:tel-num}) or differential phases can already tell a lot about the geometry and symmetry of the emission regions which then can be compared to models. The latter two modes have been employed in the previous generation of ground-based infrared interferometers like MIDI or AMBER at the VLTI. We are actively working on such issues of operational modes. For the drift mode, we started by just considering passive drifts where the pathways in the (u,v) plane approach straight lines. This results in peculiar shapes for the synthesized beam and calls for attention when applying the common inversion and deconvolution techniques like CLEAN. We think that the application of different approaches for such tasks has to be studied, for instance by invoking sparse-sampling algorithms, as has been done already for radio data with LOFAR \citep{2015A&A...575A..90G}. Too symmetric drift configurations (in terms of relative angles of the satellite paths) and too similar drift velocities will lead to degeneracies in the (u,v) plane coverage. To better fill the (u,v) plane, the application of acceleration maneuvers during a measurement campaign is necessary, which of course has an impact on the formation dynamics, on the operations as well as on the fuel budget. We plan to address these issues for all the three cases mentioned above in a future publication (Buinhas et al., in prep.).

\subsection{System architecture and stability requirements}
Based on the main requirements of the mission and the related stability requirements, a preliminary spacecraft design has been developed (see Fig.~\ref{Fig:IRASSI-sketch}), including a description of the main subsystems of the spacecraft \citep{Ferrer-Gil_2016}. Included are first estimates of the mass distribution and the power consumption budget of the different subsystems (mechanical \& thermal,  payload, attitude and orbit control, propulsion, communication, command and data handling, power). We estimate that total power of $\approx 1780$ W will be needed per satellite element, including a 20\% margin. Assuming current technology of  multi-junction GaAs-based solar cells, it is estimated that they will be able to generate 251 W/m$^2$ at the Lagrange point L2 at the end of their life. We conclude that ~8.4 m$^2$ of solar cells will be required at each spacecraft to provide the required power, which is somewhat smaller than the solar panel area for the Herschel satellite ($\approx 12$ m$^2$). The heaviest single parts within the satellite will be the primary mirror ($\approx 231$ kg) and an optical bench ($\approx 236$ kg). 
The presented spacecraft design has been used to assess the effect of environmental thermal fluctuations on the stability of the spacecraft.  The major components, except the service module, will be subject to the passive cooling by the heat shields. For the receiver/mixer components in the signal chain we will need active cooling to reach the very low temperatures of down to 4~K. Hence, these components have to be encapsulated in a cryostat. Sorption coolers could be a choice since they do not introduce excessive vibrations into the satellite system. We find that the primary mirror can introduce relatively high distortions in the critical parts of the spacecraft. This is of course due to its quite large diameter of 3.5 m, so that the mirror thermally influences its surrounding elements, including the optical bench. Based on these results it is recommended to investigate techniques to thermally isolate the primary mirror in order to reduce its influence to the dimensional stability of the spacecraft. This could be achieved by a proper mounting and protecting the primary mirror with a baffle. The optical bench probably has to be based on low-expansion-index materials like Zerodur or Silicon Carbide. It has to be further investigated whether such design features can ensure that the path between the laser metrology terminals and the science instruments remains under control. It may be necessary to consider an internal metrology setup to calibrate these internal pathways. Important conceptual work in that regard had already been done in the context of preparing for the interferometric astrometry mission SIM  \citep[cf.][]{2004SPIE.5491.1020D}. Careful alignment and thermal monitoring is of course a major point in the design of all space-borne astronomical satellites \cite[see e.g.,][for the James Webb Space Telescope]{2012SPIE.8442E..3IB}. Another important conclusion is that attitude changes can produce higher dimensional distortions than solar flux changes due to the more gradual Sun distance changes. This suggests that observations of the infrared sources should be scheduled trying to minimize the necessary orientation change of the telescope in order to minimize the introduced distortions. Adequate procedures should be developed to monitor and calibrate these effects \citep{Ferrer-Gil_2016}.

\subsection{Precise Metrology -- Optical Frequency Combs}\label{Subsection:frequency-combs}
\begin{figure}[t]
\includegraphics{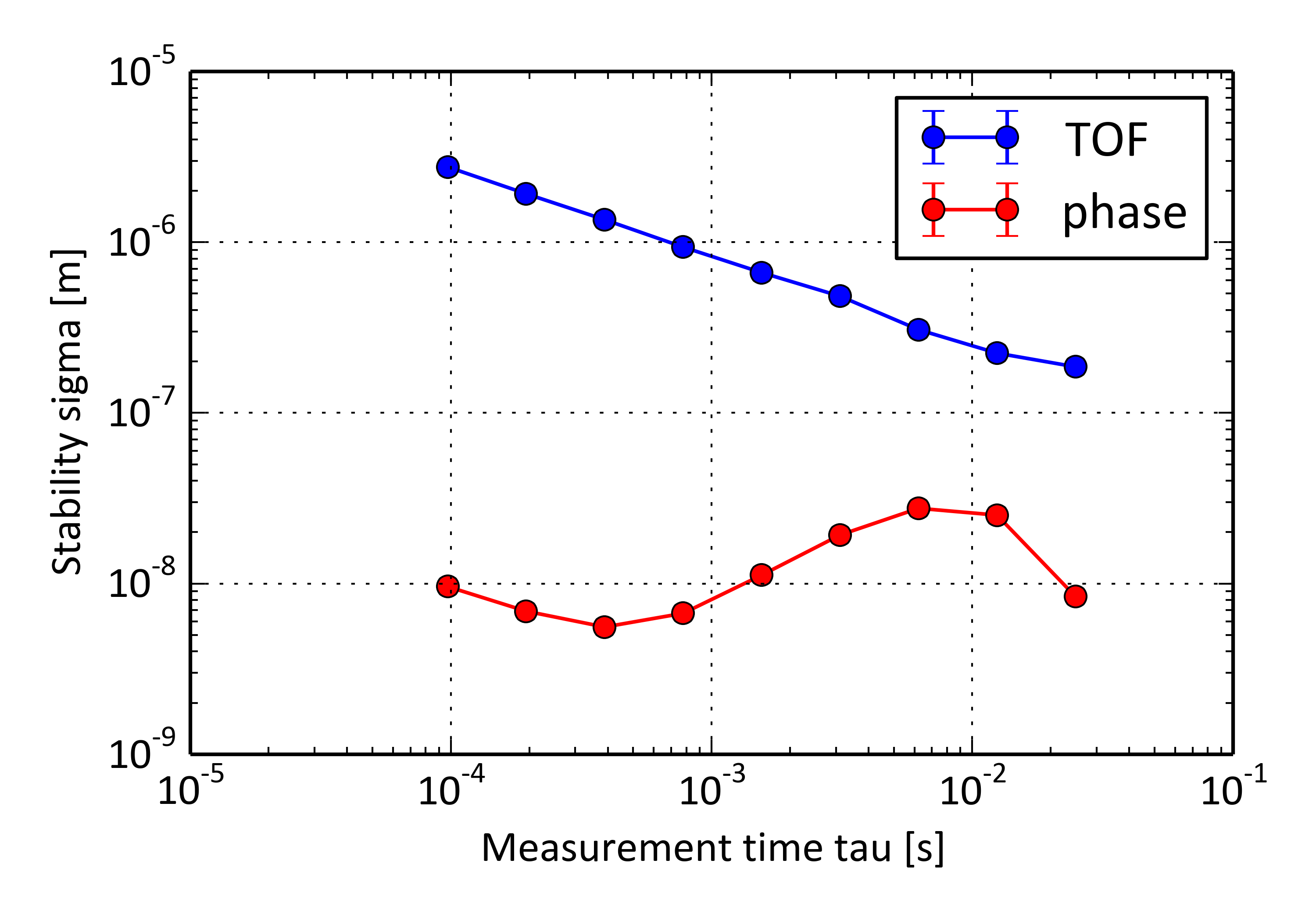}
\caption{Performance of the experimental two-combs setup for distance metrology. The upper curve is for time-of-flight (TOF) measurements and the lower curve represents the interferometric measurement. The interferometric measurement becomes unambiguous when the TOF measurement becomes $<\lambda/2$, i.e., at around 10 ms.}\label{Fig:combs}
\end{figure}
One advantage of a heterodyne system compared to direct detection inferometers is that precise distances between the satellite elements do not have to be controlled during the measurements. One needs, however, a versatile inter-satellite metrology system in order to measure the distances continuously with high time cadence. This is especially true for our particular concept of freely drifting satellite elements, for which relative motions in the order of cm/s are envisioned. We propose to employ optical frequency combs for such a metrology task. For such a comb, which is generated by periodic pulses of a femto-second laser, all resulting optical-frequency teeth have the same frequency offset to their immediate neighbours. This frequency spacing is given by the pulse repetition rate  $f_{\rm r}$ of the laser.  Typical values for $f_{\rm r}$ lie at 10 MHz up to 10 GHz. The comb is shifted with respect to zero by the offset frequency $f_{\rm 0}$. All mode frequencies of the optical frequency comb can be  transformed into radio frequencies (RF) by the comb equation $f_n = n \, f_{\rm r} + f_{\rm 0}$ and hence can be heterodyne transformed and processed using established RF electronics.  
Menlo Systems has carefully evaluated several methods how a laser frequency comb can be used for precision distance ranging. For the purpose of IRASSI, we find that double-comb interferometry \citep{2009NaPho...3..351C} offers great perspectives. For such a method, a pair of two combs, having pulse trains of slightly different repetition periods ($T_{\rm r}$ and $T_{\rm r} - \Delta T_{\rm r}$), are being used. One comb sends out the metrology signal that is reflected at the distant satellite and returns back where it is heterodyne-combined with the second comb that serves as a local oscillator, providing shotnoise-limited performance.  The concept enables us to achieve high measurement sampling that is governed by the ratio $T_{\rm r}^2 / \Delta T_{\rm r}$.  Our IRASSI system allows a real-time reduction of the data with a high duty cycle of up to 50 kHz. A customised digitiser with FPGA unit is used for data acquisition and real-time evaluation, programmed by specialists at Menlo Systems. In Fig.~\ref{Fig:combs}, we show a result from our lab measurements. Different approaches have been tested, whereby the combs were locked onto RF and onto optical references. Significantly better results have been achieved with the optical lock, providing accuracies down to $10^{-8}$ m within 10 ms. An additional advantage of the optical lock compared to RF lock is, that it can be compared and calibrated to an atomic transition.
We may come back to the question raised in Section \ref{Subsection:baselines} concerning whether also higher precision can be reached to calibrate the baseline lengths (in realtime) as is done for the antenna positions on ground for ground-based interferometers (using dedicated calibration runs in the latter case). We think that this is possible as shown by the lower measurement curve in Fig.~\ref{Fig:combs}. After 10 ms of averaging time, this particular setup resulted in around 0.01 $\mu$m accuracy. This would push the phase uncertainty of the incoming wavefronts due to baseline uncertainties to $<0.1^\circ$ even for our shortest wavelength of 50 $\mu$m. 
Currently, the lab setup is contained in a larger electronics rack. However, for the ongoing second study phase, we plan to come up with a miniaturised bread-board design. For sounding rocket projects, Menlo Systems has developed devices where the optical unit is around 1 litre in volume, and consumes less than 50 W, already fulfilling many of the requirements for a satellite application. This design will be adapted for the dual-comb ranging. We envision to have the actual frequency comb units close to the service module, where the power can be dissipated under controled conditions. The laser signals will be conducted to the terminals of the ranging system via fibers. The direction along which the terminals are mounted is not fixed and they are vectorizable within $\pm 90^\circ$ both in azimuth and elevation (relative to the optical bench frame). Furthermore, because only a small spectral section of the comb with a bandwidth of less than 1 nm is actually used for each baseline, one double comb per satellite is sufficient to support all 4 optical links to its satellite neighbors. The baseline links between the satellites will also be bi-directional, and can additionally be used for fast data transfer and exchange between the satellites. All frequency combs will be coherently locked to the same optical master reference, which will also be distributed via the optical links. Because the satellites are drifting the optical reference has to be Doppler-compensated for each satellite.

\subsection{Studies on Attitude Estimation and Control}
\subsubsection{High Accuracy ``Attitude Determination/Estimation and Control System (ADCS)'' }\label{Subsection:ADCS}
\begin{table}[ht]
\begin{tabular}{l|l|l}
\hline
Spacecraft System   & Accuracy achieved [arcsec]    & Total accuracy required \\
\hline
\hline
Attitude estimation  & 0.03 (best-case), 0.1 (worst)  & 0.4 \\
Attitude control       & 0.02 (best-case)                  &  \\
\hline      
\end{tabular}
\caption{IRASSI spacecraft pointing accuracies achieved by attitude estimation and control systems.}\label{Table:accuracies}
\end{table}
For the Herschel FIR satellite, the pointing performance, both in APE\footnote{The absolute pointing error (APE) is defined as the angular separation between the desired pointing direction and the instantaneous actual pointing direction.} and RPE\footnote{The relative pointing error (RPE) is the angular separation between the instantaneous pointing direction and the short-time average (e.g., 60 s) pointing direction at a given time period.}, could be brought to an rms of 0.9 arcsec \citep{2014ExA....37..453S}. This was eventually achieved with a robust concept of commercial optical CCD star trackers in combination with a gyro system, after some initial problems had been understood and resolved as the active mission went along. Considering our goal of 0.2 arcsec for the IRASSI antennas, this shows, that it will not be sufficient to just replicate the Herschel attitude estimation and control concept for the IRASSI antennas. 
\begin{figure}[t]
\includegraphics[width=\textwidth]{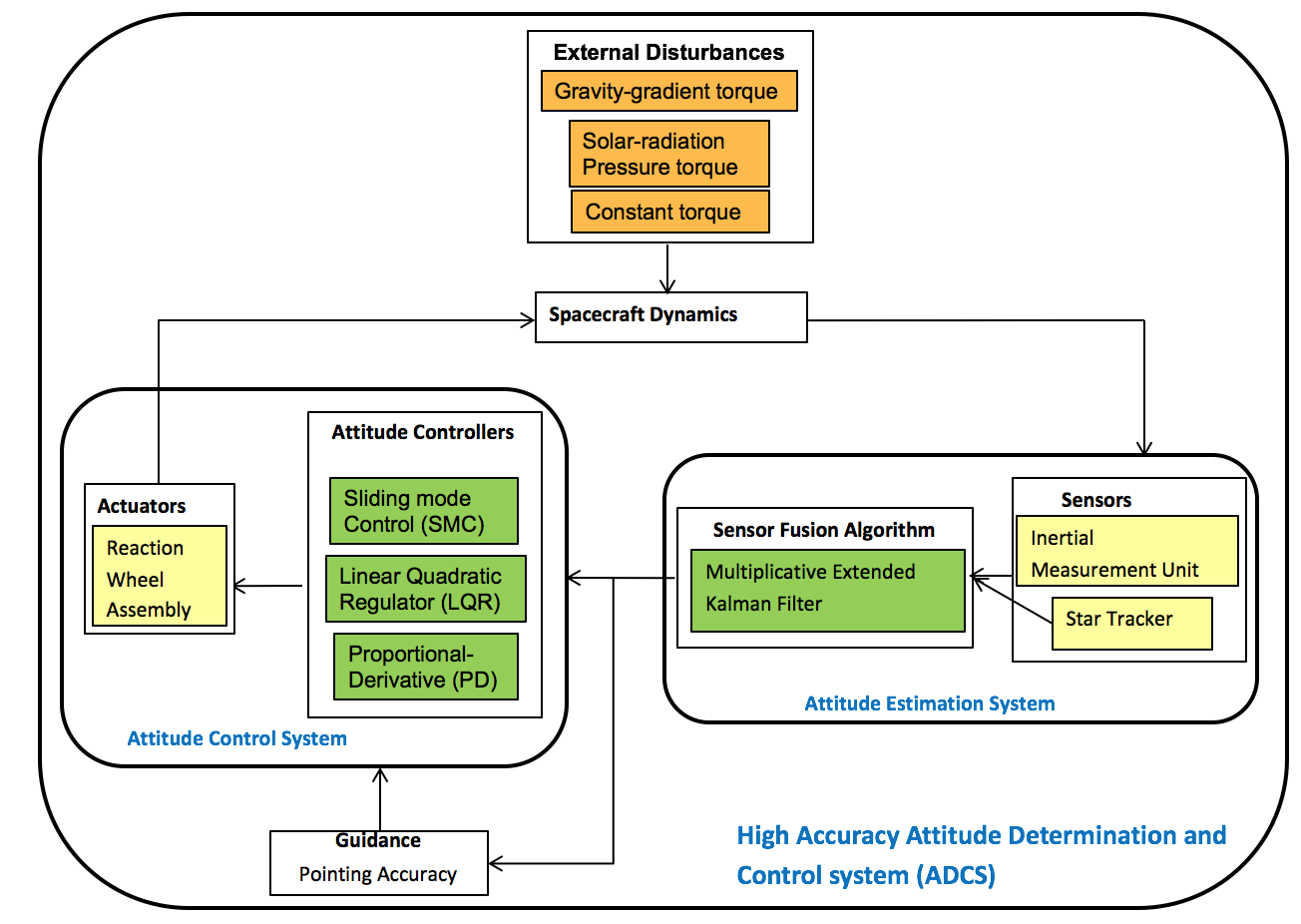}
\caption{IRASSI’s High Accuracy Attitude Determination and Control System (ADCS).}\label{Fig:ADCS}
\end{figure}
\begin{figure}[t]
\begin{center}
\includegraphics[width=0.5\textwidth]{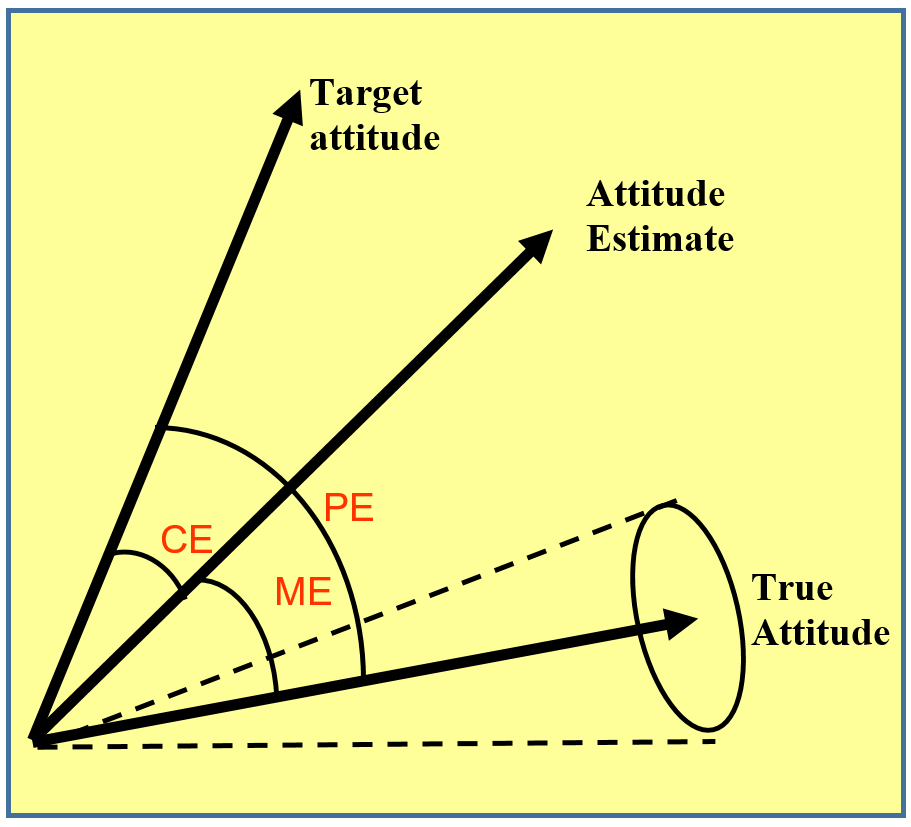}
\end{center}
\caption{Attitude definitions: Pointing Error (PE), Measurement Error (ME) and Control Error (CE).}\label{Fig:Attitude-Definitions}
\end{figure}
Therefore, the IRASSI study covers active research on attitude determination/estimation and related control systems.  Figure \ref{Fig:ADCS} shows the components of IRASSI’s high accuracy ``Attitude Determination and Control System (ADCS)'' which are Attitude Estimation System (AES) and Attitude Control System (ACS) that were designed and developed, and according to our simulations achieve unprecedented stringent attitude estimation and control (best-case) accuracies of 0.03 and 0.02 arcsec respectively (cf. Table \ref{Table:accuracies}).  As shown in Fig.~\ref{Fig:Attitude-Definitions}, the total pointing accuracy/error can be separated into two parts, i.e. pointing accuracy from AES and ACS systems known as Measurement Error (ME) and Control Error (CE), respectively. This leads to (theoretically) achieving the imposed total pointing accuracy requirement of 0.4 arcsec (with the goal of 0.2 arcsec) on the IRASSI spacecrafts. This also leaves enough margin for pointing errors due to other space  sub-systems like IRASSI reaction wheels (required for attitude slews and part of ongoing IRASSI work), thruster firing (used for IRASSI formation control), cryogenic coolers, propellant sloshing and other structural instabilities, to name a few.

\subsubsection{Attitude Estimation System (AES)}
\begin{figure}
\includegraphics[width=\textwidth]{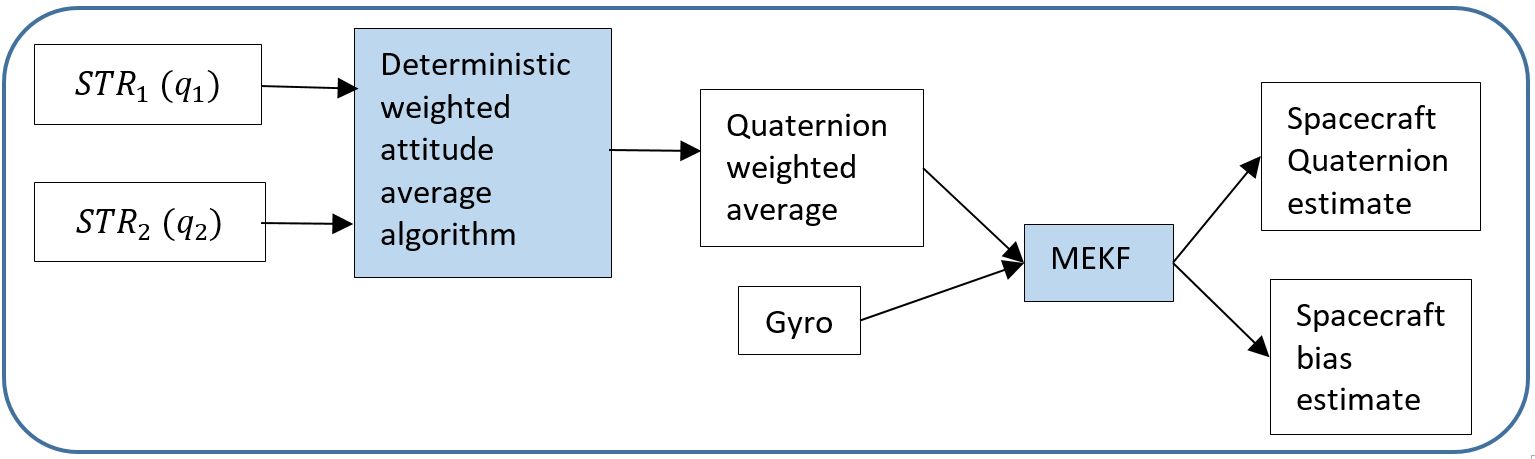}
\includegraphics[width=\textwidth]{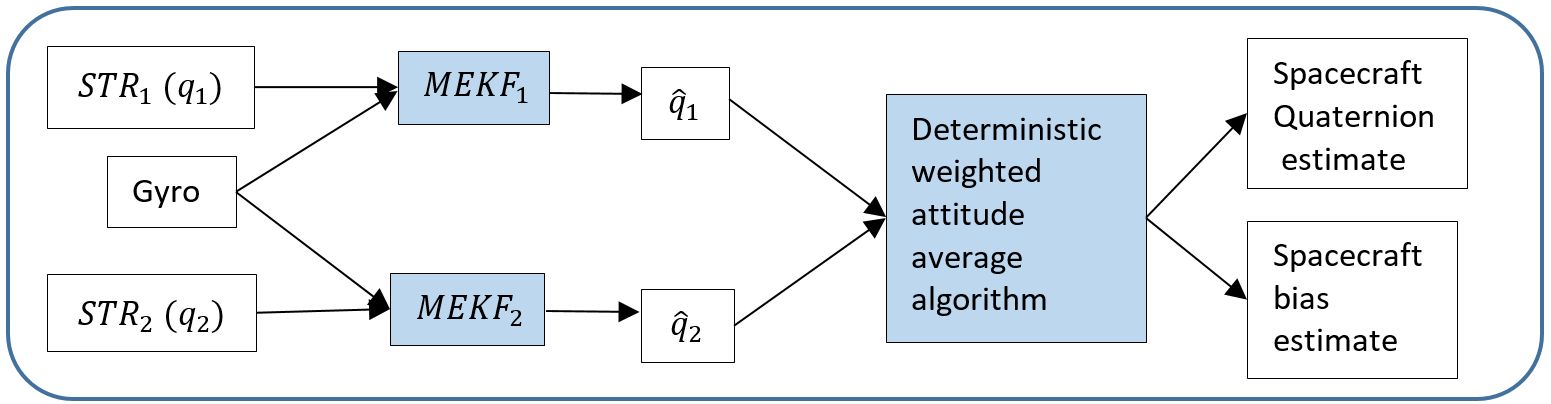}
\caption{Centralized (upper) and de-centralised (lower) filtering schemes.}\label{Fig:schemes}
\end{figure}

\begin{figure}
\begin{center}
\includegraphics[width=0.8\textwidth]{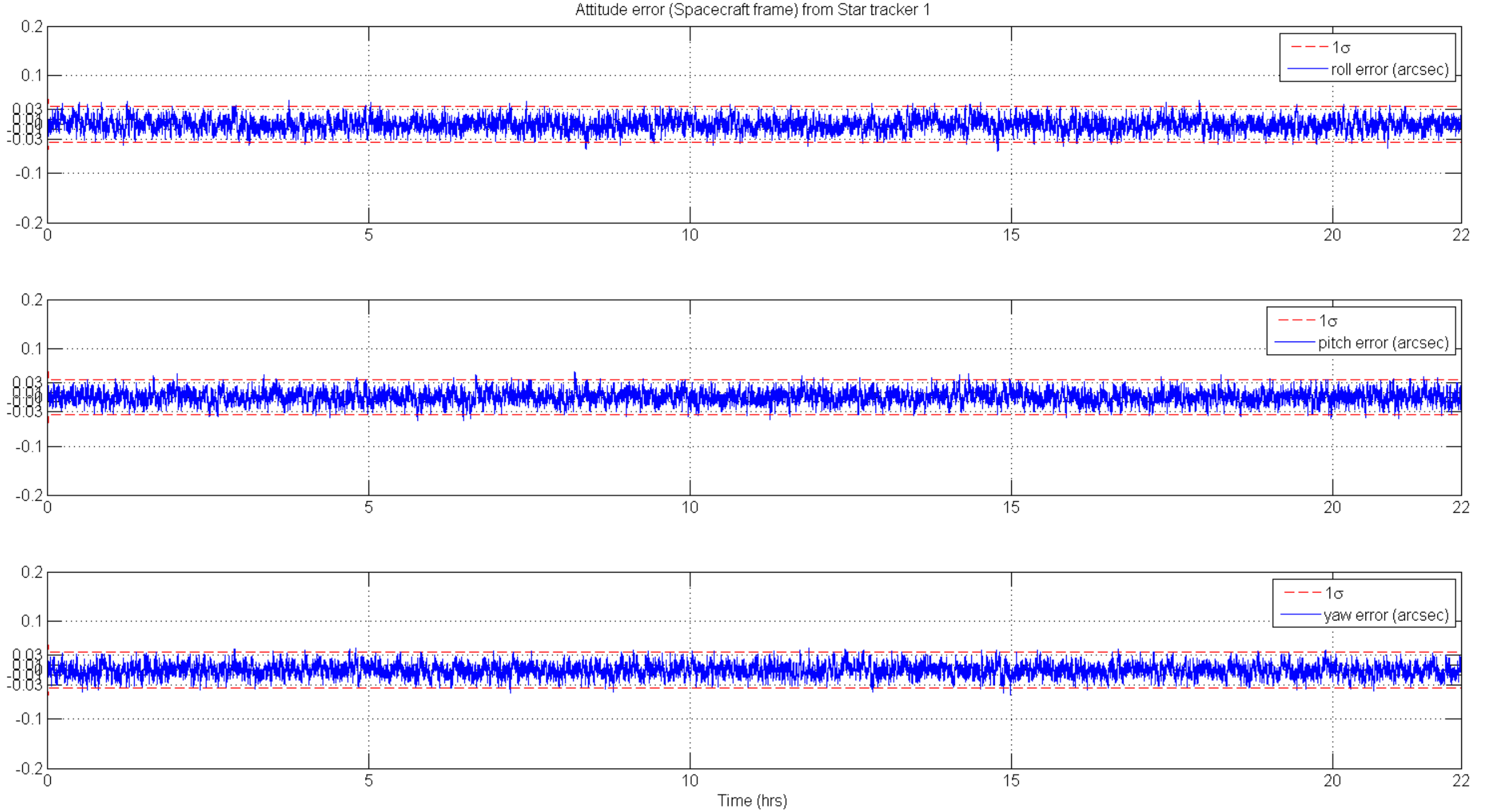}
\includegraphics[width=0.8\textwidth]{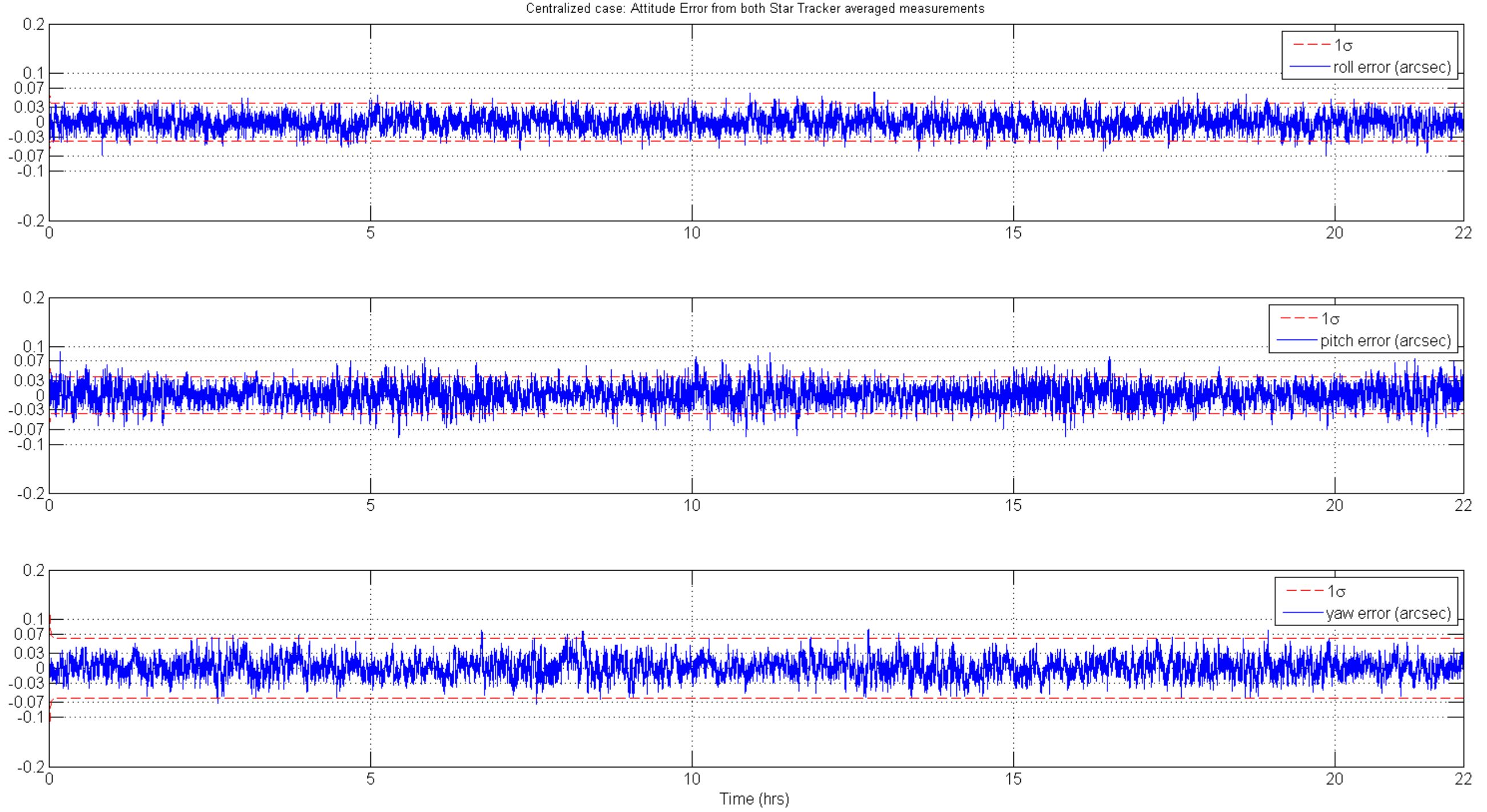}
\includegraphics[width=0.8\textwidth]{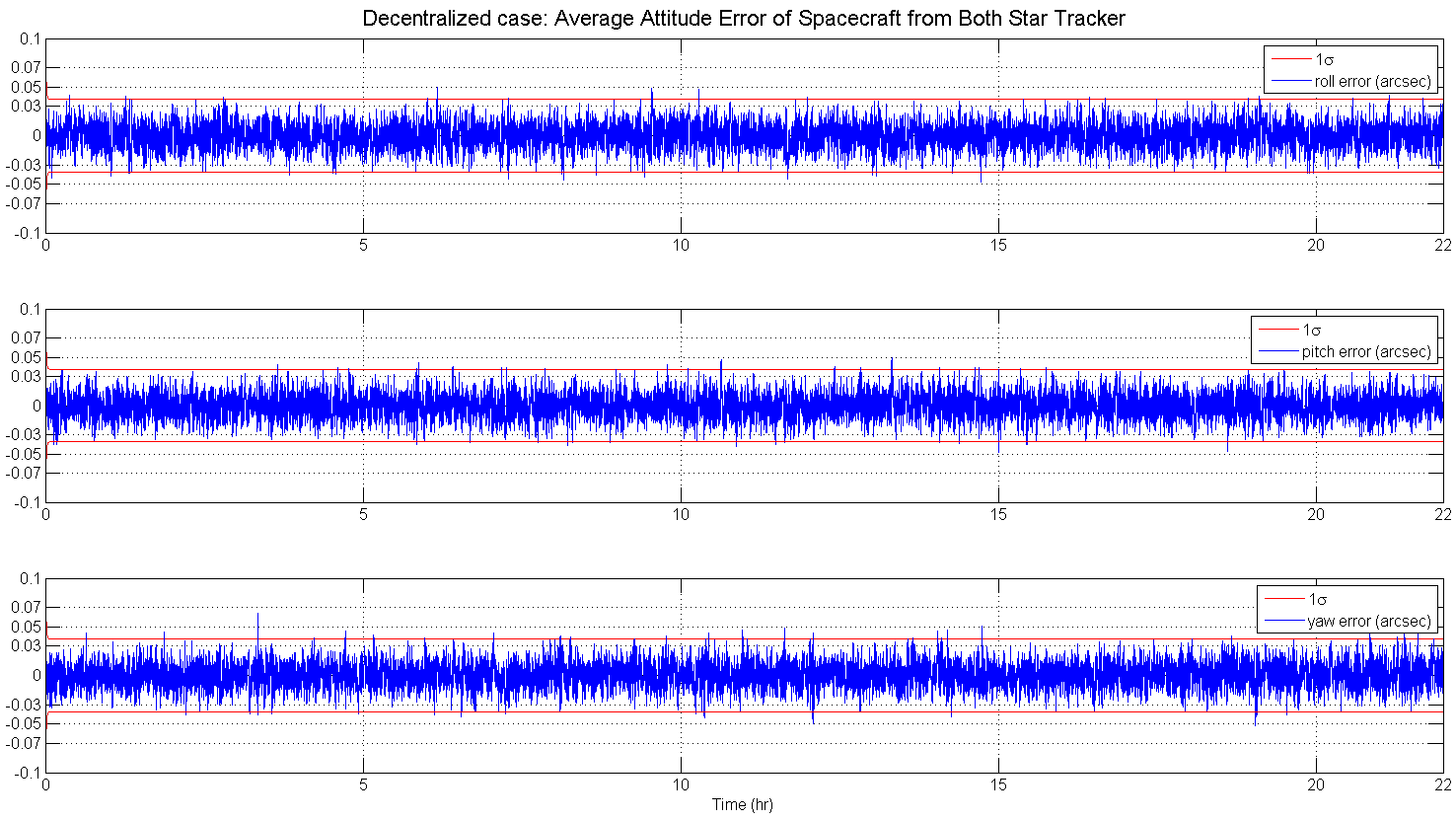}
\end{center}
\vspace*{-0.5cm}
\caption{The attitude accuracy (in arcsec) found in our simulations for a 22-hour interval for three cases. Up: one star tracker and Gyro fusion; Middle: centralised case; Bottom: decentralised case.}\label{Fig:sim-results}
\end{figure}

\begin{table}
\includegraphics[width=\textwidth]{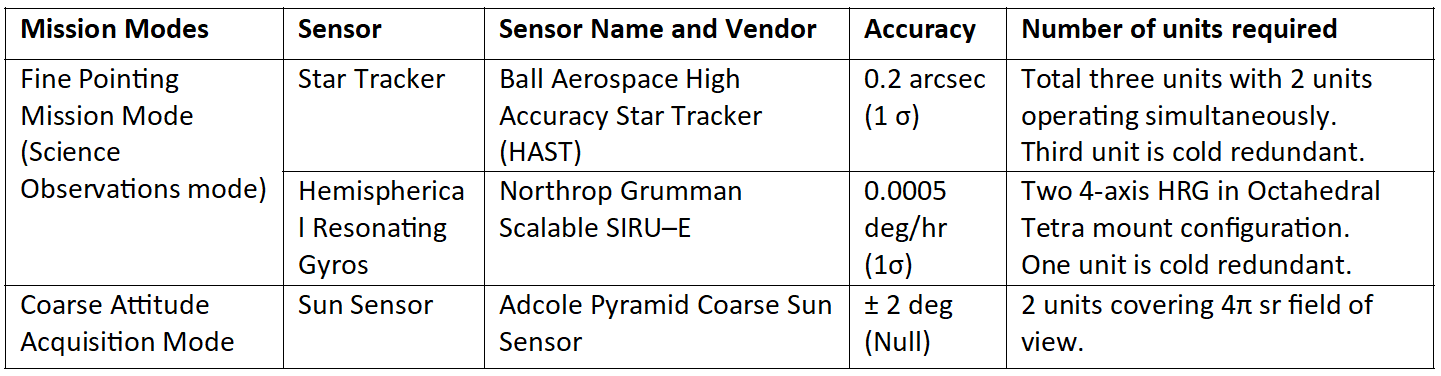}
\caption{Sensor Suite for High accuracy Attitude Estimation System.}\label{Table:sensor-suite}
\end{table}
High Accuracy attitude estimation accuracy of 0.03 arcsec is accomplished by AES which involves carefully selected high accuracy commercial-off-the shelf sensors and implementing a state-of- the-art optimal attitude estimation algorithm for sensor data fusion in Matlab. Table~\ref{Table:sensor-suite} shows the sensor-suite selected for high accuracy AES. During the fine-pointing/scientific observation mission mode of IRASSI, data from a star tracker (single or multiple) is fused with data from a gyroscope assembly (where each assembly consists of 4-gyro heads mounted in an octahedral tetrad configuration) via a Multiplicative Extended Kalman Filter (MEKF). For coarse attitude acquisition mode, measurements from a Sun Sensor and Gyroscopes are fused via MEKF. Three filtering schemes are developed for fine-pointing mode (as shown in Fig.~\ref{Fig:schemes} and Table~\ref{Table:mission-modes}) where a 3-axis gyroscope is fused with single and two simultaneously operating (oriented in orthogonal directions) star trackers in Centralized and Decentralized Schemes. Table \ref{Table:(de)centralised} lists the attitude accuracies achieved by simulations (as shown in Fig.~\ref{Fig:sim-results}) of these above mentioned schemes, all of which utilize an analytic filter algorithm named MEKF as explained above.
IRASSI is the first mission that makes the information on fusion of two simultaneously operating star trackers for attitude estimation public, even though this idea is also used for the BepiColombo mission to Mercury. This study also led to the placement (position and orientation) of the IRASSI sensors onboard IRASSI spacecraft. This is also indicated in Fig.~\ref{Fig:IRASSI-sketch}. More details can be found in \citet{Bhatia_2017}. 
\begin{table}
\begin{tabular}{l|l}
\hline
Estimation (Sensor Fusion) Cases   & Accuracy achieved  \\
\hline
\hline
1 STR and Gyro fusion  &  0.03 arcsec (1$\sigma$) \\
Centralized case           &  0.03 arcsec (1$\sigma$) \\
Decentralized case        &  0.03 arcsec (1$\sigma$)  \\
\hline 
\end{tabular}
\caption{Pointing accuracy results for Centralized and Decentralized cases}\label{Table:(de)centralised}
\end{table}

\subsubsection{High Accuracy Attitude Control System (ACS)}
Feasibility check of three attitude control algorithms, namely Sliding Mode control (SMC), Linear Quadratic Regulator (LQR) and Proportional-Derivative (PD) control led to an achievement of 0.02 arcsec for small-angle rest-to-rest maneuvers. The high accuracy pointing requirement that IRASSI imposes on the ADCS is for ``small angle maneuvers'' because once the science observations start, i.e., once the spacecraft points to a scientific target, there will be only small deviations in the spacecraft attitude from this scientific target.  But since ``large angle maneuvers'' (that are required by many other telescopic and non-telescopic space missions) are the most difficult attitude maneuvers to perform due to the nonlinearities of the system dynamics, properties of attitude parameters utilized, accumulation of disturbances and exposure to varying disturbances, we check the feasibility of these control algorithms in achieving tens-of-subarcsec accuracy for these maneuvers, too. Three disturbances that were fed-forward to the spacecraft nonlinear dynamics, namely gravity-gradient torque, solar-radiation pressure torque and a constant torque (to account for unforeseen disturbances, shown in Fig.~\ref{Fig:ADCS}, left) are that encountered by IRASSI spacecraft in its halo-orbit at Sun-Earth L2. \\
Stability analysis of these controllers showed their robustness regarding these disturbances as shown in Table~\ref{Table:rest-to-rest}. These controllers are “almost” globally stable. Quaternions which are chosen as the attitude parameters for ADCS, have an ambiguity when describing rotations within 360 degrees; both, a certain quaternion and its inverse describe the same rotation within the Special Orthogonal, SO(3), space. Hence due to the ambiguity of quaternions within 360 degree, the stability of these controllers is relaxed to “almost” global asymptotic stability in the sense of Lyapunov where asymptotic stability is defined over an “open” and dense set in SO(3). 
All three control algorithms are feasible for IRASSI’s fine-pointing and coarse attitude-acquisition modes as shown in Table \ref{Table:mission-modes} and \ref{Table:rest-to-rest}. These results are substantiated by simulations of these analytical control algorithms. More details about this study can be found in Bhatia, D. in prep. (``Final Report: IRASSI 2 Phase 1'', to be published at Hanover University Library).  \\
\begin{table}[t]
\begin{tabular}{l|l|l}
\hline
IRASSI Mission Modes   & Attitude Estimation filters   &  Corresponding  \\
                                 &                                     &  Attitude Control  \\
\hline
\hline
Fine-pointing mode      &  Gyro + 1 Star Tracker        &    Sliding Mode  \\
                                &                                       &  Controller (SMC), \\
                                &  Gyro + 2 Star Trackers       & Linear Quadratic \\
                                &  (centralised)                     & Regulator (LQR),  \\
                                &  Gyro + 2 Star Trackers      & Linear \& nonlinear  \\
                                &  (de-centralised)                &  PD controllers\\
\cline{1-2} 
Coarse Attitude  & Gyro + coarse Sun sensor           & \\
Acquisition Mode &                                             & \\
\hline
\end{tabular}
\caption{IRASSI mission modes and the corresponding attitude filters and controllers.}\label{Table:mission-modes}
\end{table}
In summary, 0.03 arcsec and 0.02 arcsec accuracies from attitude estimation and control systems, respectively, {\rm lead} to a total of 0.05 arcsec pointing accuracy from ADCS (theoretically). Hence, we conclude that IRASSI’s desired goal accuracy of 0.2 arcsec can be achieved, leaving sufficient margins for other factors that affect the pointing accuracies as discussed before. Currently, we are investigating the reaction wheel assembly required for physically rotating IRASSI spacecraft, specifically during fine pointing mode. Also control allocation algorithms where the control torques calculated by the attitude control algorithms are distributed amongst the redundant reaction wheels to physically maneuver IRASSI telescope are under investigation, which will provide a realistic insight into the performance of the complete ACS. 
\begin{table}
\includegraphics[width=\textwidth]{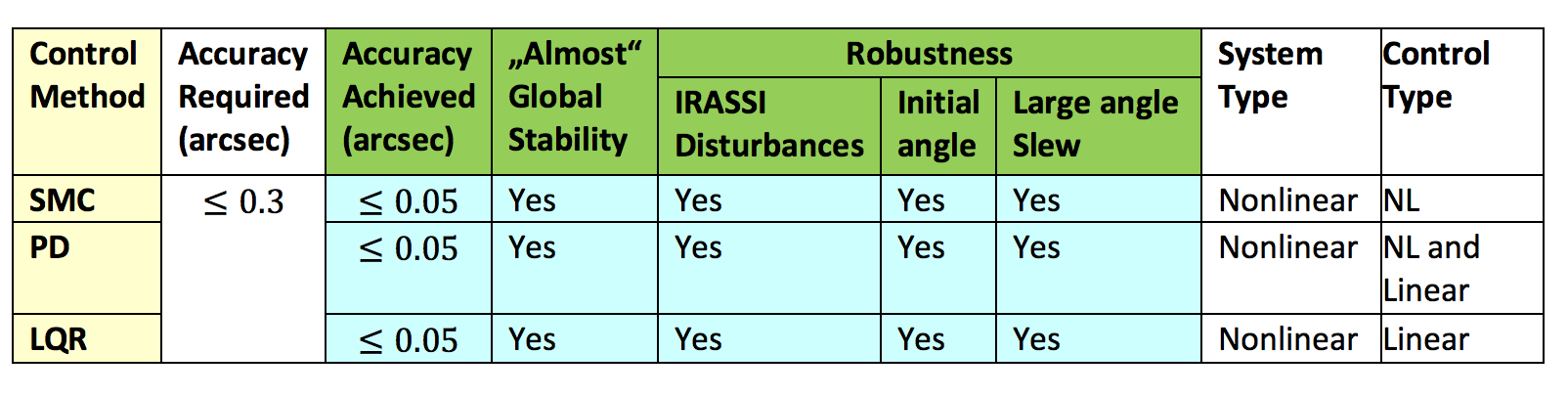}
\caption{Performance Comparison of SMC, PD and LQR control design for IRASSI Spacecraft attitude tracking rest-to-rest maneuvers.}\label{Table:rest-to-rest}
\end{table}

\subsection{Preliminary sensitivitiy estimates}\label{Subsection:sensitivity}
\begin{figure}[t]
\begin{center}
\includegraphics[width=0.7\textwidth]{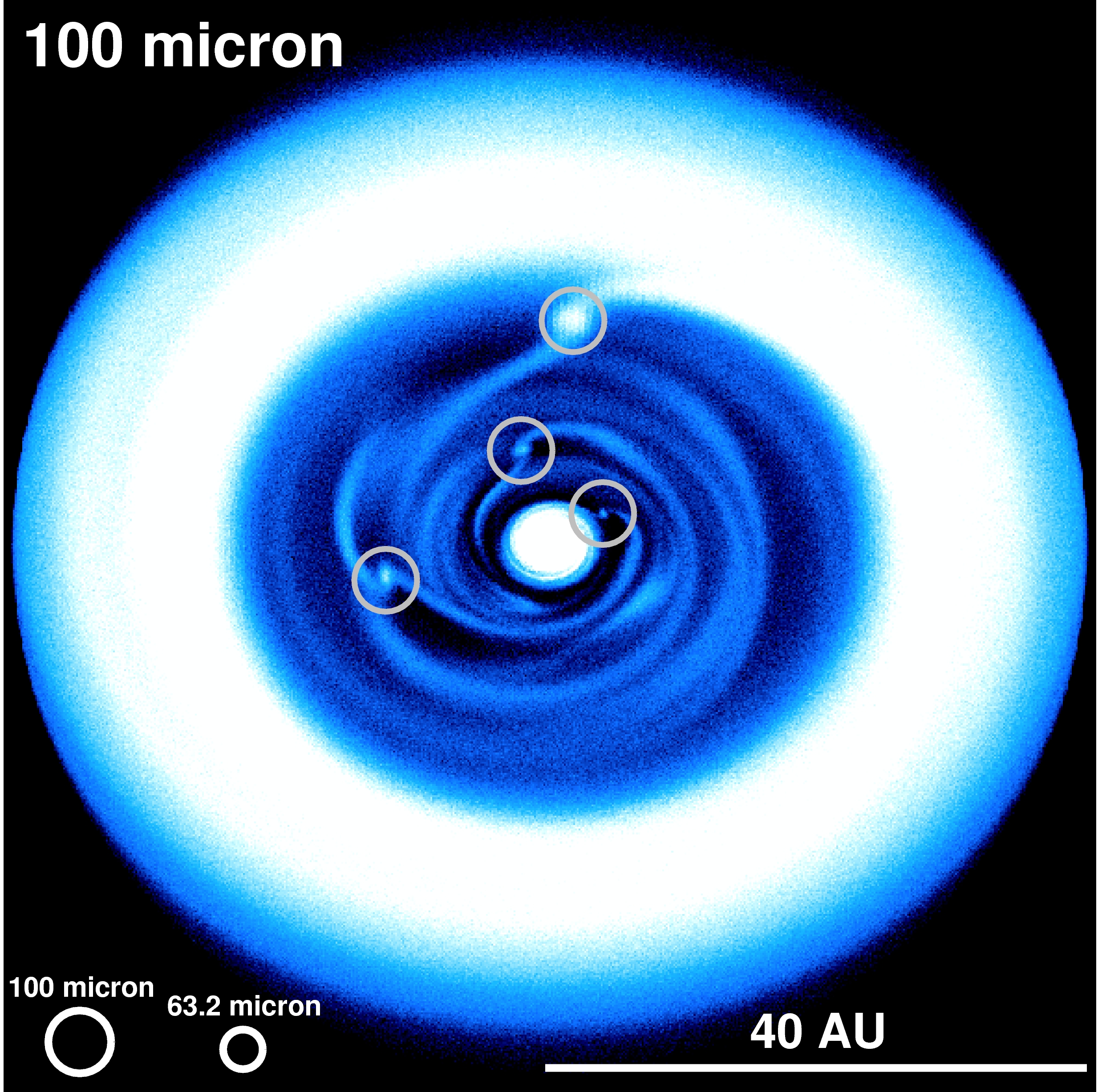}
\end{center}
\caption{Radiative transfer simulation at a wavelength of 100 $\mu$m on top of a hydro simulation of a transitional disk with four embedded proto-planets. Resolution elements for IRASSI at 100 $\mu$m and 63.2 $\mu$m (the wavelength of the strong [OI] fine-structure line) are indicated. The underlying simulation data have first been shown in \citet{2014SPIE.9146E..11K}.}\label{Fig:disk}
\end{figure}

One future task will be to include many aspects of the IRASSI design in a comprehensive performance simulation. Here, we want to give just an order of magnitude estimate for the achievable sensitivity, to guide the reader what to expect from a facility like IRASSI. \\
We will employ the standard sensitivity equation for radio interferometry. We use the following form:
\begin{equation}\label{Equ:sensitivity}
\sigma = \frac{2 {\rm k_B} \,\,\, T_{\rm sys}}{\eta_{\rm corr} \,\,\, \nicefrac{\pi}{4}\, D_{\rm Tel}^2 \,\,\, \eta_{\rm ap} \,\,\, \sqrt{N_{\rm A} (N_{\rm A} - 1) \,\,\, N_{\rm pol} \,\,\, \Delta \nu \,\,\, \tau}}
\end{equation}
$\sigma$ --  noise in the correlated data, in W m$^{-2}$ Hz$^{-1}$ \\
k$_{\rm B}$ -- Boltzmann constant \\
$T_{\rm sys}$ -- system temperature, in K \\
$\eta_{\rm corr}$ -- correlator efficiency \\
$\eta_{\rm ap}$ -- aperture efficiency \\
$D_{\rm Tel}$ -- single-dish telescope diameter; 3.5 m for IRASSI \\
$N_{\rm A}$ -- number of antennas; 5 for IRASSI \\
$N_{\rm pol}$ -- number of polarisation products, 2 for IRASSI  \\
$\Delta \nu$ -- frequency bandwidth, in Hz \\
$\tau$ -- on-source integration time, in s \\\smallskip\\
For simplicity, we assume here that $T_{\rm sys}$ is dominated by the receiver/mixer noise temperatures. For $\eta_{\rm corr}$ we adopt the value of 0.88, a number that is commonly invoked for correlators processing input stream data with 4-level 2-bit digitisation \citep[e.g.,][]{1989ASPC....6...59D}. For the aperture efficiency $\eta_{\rm ap}$ we have no firm values yet. Looking back to the Herschel satellite, its telescope (with the same 3.5 m diameter as anticipated for IRASSI) had an aperture efficiency on the order of 0.6, as measured by HIFI calibration observations\footnote{https://www.cosmos.esa.int/documents/12133/996743/The+HIFI+Beam+-+Release+No1+Release+Note+for+Astronomers}. We therefore adopt this value as a conservative preliminary estimate.  \\
We give two numbers as examples. In both cases we assume that a $T_{\rm sys}$ on the order of three times the theoretical quantum limit (TQL $\approx$ 48 K/THz) can be achieved; admittedly a very ambitious goal, but in line with recommendations in \citet{2017arXiv170100366R}. One estimate is relevant for line measurements, e.g., at the frequency of the [CII] line at 1.9 THz. A 1 km/s channel at this frequency covers 6.333 MHz. After an integration time of 20 hours (the typical time for IRASSI to let the drifting satellites fill the (u,v) plane), the rms noise in such a channel would amount to 35 mJy. For a line with 3 km/s FWHM line width, and sampled with 1 km/s channels, this would be equivalent to a 5-$\sigma$ line flux of $3.5 \times 10^{-20}$ W/m$^2$.  The second computation is done for the frequency of 3 THz, corresponding to 100 $\mu$m wavelength. Assuming that the receiver/backend system delivers a full bandwidth of 16 GHz in dual polarisation, after 20 hours of integration a continuum noise value of 1.1 mJy can be achieved. (This scales to 4.9 mJy for 1 hour of integration.)
To give some perspective, we include here a simulation of a pre-transitional protoplanetary disk, originally published in \citet{2014SPIE.9146E..11K}. This is a result of a hydrodynamics simulation for the case when four Jupiter-mass proto-planets interact at different orbital radii with the disk material. The disk is inclined by 30$^\circ$, the full field-of-view is 80 au. The subsequent radiative transfer simulates the appearance at 100 micron. The result is shown in Fig.~\ref{Fig:disk}. We mark the locations of the four planets with circles the size of an IRASSI resolution element at this wavelength. We emphasize that the outermost proto-planet concentration (the compact brightness peak 20 au from the central star) attains a flux density of around 23 mJy according to this simulation. Hence, at least regarding the sensitivities, such an object could be well detectable with IRASSI.\\
Equation \ref{Equ:sensitivity} makes it obvious that the telescope diameter $D_{\rm Tel}$ is an important factor for the sensitivity, since it enters this formula with the squared power. On the other hand, for larger numbers of telescope elements, [$N_{\rm A} (N_{\rm A} - 1)]^{0.5}$ approaches $N_{\rm A}$, hence just a linear term. Thus, it is more difficult to gain sensitivity by adding more satellite elements. To achieve the same sensitivity as a five-element interferometer with 3.5-m antennas, it would take nine satellites with 2.5 m diameter, for instance.

\section{Summary and outlook}

We have shown the principle design of IRASSI, a free-flying FIR interferometer with five satellite elements at the Lagrange point L2, which shall cover an essential frequency range between 1–6 THz and shall deliver an angular resolution of better than 0.1 arcsec over the whole working range. Compared to other concepts employing direct detection systems, IRASSI adopts the heterodyne principle for detection. As a consequence, satellite distances need not be actively controlled. They have to be measured accurately, however, with high time cadence to estimate the geometric delay terms pivotal for the raw data correlation. In IRASSI, we employ a metrology concept based on coupled laser frequency combs to handle this task. We are actively working on perfecting such a system in the lab, and to come up with a miniaturised breadboard design. \\
A preliminary spacecraft architecture has been designed, including a description of the main subsystems of the spacecraft, as well as the corresponding estimates of the mass distribution and the power consumption budget of each subsystem. Furthermore, we have worked out a full mission scenario for IRASSI. We have modelled how a swarm of five satellites is to be deployed into a Halo orbit around L2. We follow the dynamics of the swarm and identify possible 3D configurations, for which detailed sky coverage and exclusion regions have been derived for different conditions. We are actively working on how to navigate such a swarm, how the attitude of the satellites can be estimated with high precision, how the metrology distance measurements are to be incorporated in order to assess the geometric delays relevant for the correlation process of the interferometric science data, and how the overall baseline estimation procedure can deliver the required accuracy of $< 5 \,\mu$m. \\
Looking onto the technological side of the measurement and correlation process, the recent developments for HEB mixers, quantum cascade laser LO systems and new ASIC-based technology for correlators are promising. We recommend that technology pathways towards a THz mixer performance reaching 3–5 times the quantum limit should be pursued to enable sensitive measurements, in order to offer IRASSI to a wide scientific community. 
In the ongoing second project phase, we will put further emphasis on key issues that have emerged from our initial studies, especially how to model thermal and mechanical distortions accurately to integrate them in the baseline determination measurement chain, along with the ranging system and relative attitude measurements. In particular, it will be important to work out in detail how to link the laser metrology terminals to the local phase centres for the science signals. For this, a kind of local tie calibration, based on an internal metrology system, has probably to be included to assess deformations of the hardware infrastructure and other disturbances that can affect the interpretation of the inter-satellite metrology. \\

\noindent {\bf Acknowledgements} 
IRASSI is a project financed by the German Ministry of Economy and Energy and administered by the German Aerospace Center, Space Administration (DLR, Deutsches Zentrum für Luft-und Raumfahrt, FKZ 50NA1325-28, 50NA1713-16, 50NA1815-18). This work is a cooperation between the Institute of Space Technology \& Space Applications of the Bundeswehr University (Munich, Germany), the Max Planck Institute for Astronomy (Heidelberg, Germany), the Institute of Flight Guidance from the Technische Universität Braunschweig (Braunschweig, Germany) and Menlo Systems GmbH (Martinsried, Germany) as an industry partner. We thank Stefan Kraus for providing us with the 100 $\mu$m simulation of the protoplanetary disk shown in Fig.~\ref{Fig:disk}. These simulations are by courtesy of  Ruobing Dong, Barbara Whitney, and Zhaohuan Zhu.



\bibliographystyle{elsarticle-harv} 
\bibliography{Linz_IRASSI}





\end{document}